\documentclass[aps,prd,preprintnumbers,groupedaddress,nofootinbib,amssymb,notitlepage,eqsecnum]{revtex4-1}
\usepackage{here}
\usepackage{graphicx}
\usepackage{amsmath,amsthm,amssymb}
\usepackage{bm}
\usepackage{color}

\usepackage{amsfonts}
\usepackage{dcolumn}
\usepackage{hyperref}
\allowdisplaybreaks[1]
\usepackage{stackengine}

\begin{document}
\newcommand{\newc}{\newcommand}

\newcommand{\rk}{\textcolor{red}}
\newc{\be}{\begin{equation}}
\newc{\ee}{\end{equation}}
\newc{\ba}{\begin{eqnarray}}
\newc{\ea}{\end{eqnarray}}
\newc{\D}{\partial}
\newc{\rH}{{\rm H}}
\newc{\rd}{{\rm d}}
\newc{\Mpl}{M_{\rm Pl}}
\newcommand{\rBH}{r_{s}}
\newcommand{\rc}{r_{c}}
\newcommand{\rh}{r_{h}}
\newcommand{\Xh}{X_{h}}
\newcommand{\hr}{\hat{r}}
\newcommand{\odd}[1]{{#1}_\chi}
\newcommand{\even}[1]{{\rm\bf #1}}
\newcommand{\ma}[1]{\textcolor{magenta}{#1}}
\newcommand{\cy}[1]{\textcolor{cyan}{#1}}
\newcommand{\mm}[1]{\textcolor{red}{#1}}

\preprint{YITP-22-06, WUCG-22-01}

\title{Linear stability of black holes in shift-symmetric Horndeski theories\texorpdfstring{\\}{}
with a time-independent scalar field}

\author{Masato Minamitsuji,$^{1}$ Kazufumi Takahashi,$^{2}$ 
and Shinji Tsujikawa$^{3}$}

\affiliation{
$^{1}$Centro de Astrof\'{\i}sica e Gravita\c c\~ao - CENTRA, Departamento de F\'{\i}sica, Instituto Superior T\'ecnico - IST, Universidade de Lisboa - UL, Av.~Rovisco Pais 1, 1049-001 Lisboa, Portugal\\
$^2$Center for Gravitational Physics, Yukawa Institute for Theoretical Physics, Kyoto University, 606-8502, Kyoto, Japan\\
$^3$Department of Physics, Waseda University, 3-4-1 Okubo, Shinjuku, Tokyo 169-8555, Japan}

\begin{abstract}
We study linear perturbations about static and spherically symmetric black holes with a time-independent background scalar field in shift-symmetric Horndeski theories, whose Lagrangian is characterized by coupling functions depending only on the kinetic term of the scalar field~$X$. We clarify conditions for the absence of ghosts and Laplacian instabilities along the radial and angular directions in both odd- and even-parity perturbations. For reflection-symmetric theories described by a k-essence Lagrangian and a nonminimal derivative coupling with the Ricci scalar, we show that black holes endowed with nontrivial scalar hair are unstable around the horizon in general. This includes nonasymptotically flat black holes known to exist when the nonminimal derivative coupling to the Ricci scalar is a linear function of $X$. We also investigate several black hole solutions in nonreflection-symmetric theories. For cubic Galileons with the Einstein-Hilbert term, there exists a nonasymptotically flat hairy black hole with no ghosts/Laplacian instabilities. Also, for the scalar field linearly coupled to the Gauss-Bonnet term, asymptotically flat black hole solutions constructed perturbatively with respect to a small coupling are free of ghosts/Laplacian instabilities.
\end{abstract}

\date{\today}

\maketitle

%%%%%%%%%%%%%%%%%%%%%%%%%%%%%%%%%%%%%%%%%%
\section{Introduction}
\label{introsec}
%%%%%%%%%%%%%%%%%%%%%%%%%%%%%%%%%%%%%%%%%%

Black holes (BHs) are fundamental objects whose existence 
is theoretically predicted by general relativity (GR) 
and other gravitational theories. 
With the dawn of gravitational-wave 
astronomy~\cite{LIGOScientific:2016aoc}, we can 
now probe physics of BHs and possible deviations 
from GR at strong gravity 
regimes~\cite{Berti:2015itd,Barack:2018yly,Berti:2018cxi,Berti:2018vdi}. 
The discovery of BH 
shadows~\cite{EventHorizonTelescope:2019dse} 
also opened up a new window 
for exploring the properties of BHs. 
Under this observational status, it is important 
to classify what kinds of BHs exist in 
the presence of additional degree(s) of freedom 
like a scalar field or in gravitational 
theories beyond GR.

In GR with an electromagnetic field, a uniqueness theorem 
states that asymptotically flat and stationary BH 
solutions are characterized only by 
mass, angular momentum, and electric 
charge~\cite{Israel:1967wq,Carter:1971zc,Ruffini:1971bza}.
This ``no-hair'' property of BHs also holds for 
a minimally coupled canonical scalar field~$\phi$~\cite{Hawking:1971vc,Bekenstein:1972ny},
a minimally coupled k-essence~\cite{Graham:2014mda},
as well as a scalar field nonminimally 
coupled to the Ricci scalar~$R$ in the form 
$F(\phi) R$~\cite{Hawking:1972qk,Bekenstein:1995un,Sotiriou:2011dz,Faraoni:2017ock}.
The no-hair theorem does not persist in scalar-tensor 
theories containing derivative couplings 
like $G_4(X)R$ in the Lagrangian, where 
$X=-g^{\mu\nu}\partial_{\mu}\phi \partial_{\nu}\phi/2$ is 
the kinetic term of the scalar field. 
Such derivative couplings can be accommodated in 
a framework of so-called Horndeski theories, 
which form the most general class of
scalar-tensor theories with second-order 
Euler-Lagrange equations~\cite{Horndeski,Def11,KYY,Charmousis:2011bf}.

The Lagrangian of Horndeski theories contains four 
coupling functions~$G_{2,3,4,5}$
depending on both $\phi$ and $X$.
If we impose the invariance under the constant
shift~$\phi \to \phi+c$, the functions~$G_{2,3,4,5}$ 
depend only on $X$.
In such shift-symmetric Horndeski theories, 
Hui and Nicolis~\cite{Hui:2012qt} argued that 
a no-hair result of 
BHs holds under the following three 
hypotheses~\cite{Babichev:2016rlq}: 
\begin{enumerate}
\renewcommand{\theenumi}{\roman{enumi}}
\renewcommand{\labelenumi}{(\theenumi)}
\setlength{\itemsep}{0cm}
\item 
\label{item1}
The background geometry 
is static and spherically symmetric and 
the scalar field is also static 
[see the ansatz~\eqref{BGmetric}], 
i.e., the character of the scalar field is spacelike ($X<0$).
\item 
\label{item2}
The spacetime is asymptotically flat with a 
vanishing radial field derivative~$\phi'(r) \to 0$ 
at spatial infinity ($r \to \infty$)
and the norm of 
the Noether current associated with the shift symmetry 
is finite on the BH horizon.
\item 
\label{item3}
A canonical kinetic term~$X$ is present 
in the Lagrangian and the $X$-derivatives of 
$G_{2,3,4,5}$ contain only positive or zero 
powers of $X$. 
\end{enumerate}
Namely, under these assumptions, we end up with the 
no-hair BH solution, i.e., $\phi'(r)=0$ everywhere.

If we violate at least one of the conditions given above, 
it is possible to realize hairy BH solutions 
endowed with nontrivial scalar hair.
For a scalar field with the linear dependence on time~$t$ of
the form $\phi=qt+\Phi(r)$,
which evades 
the hypothesis~(i), 
there exists a stealth Schwarzschild 
solution~\cite{Babichev:2013cya}.\footnote{The existence conditions for stealth solutions with constant $X$ in higher-order scalar-tensor theories were specified in Refs.~\cite{Motohashi:2019sen,Takahashi:2020hso}.}
If the asymptotic flatness of spacetime is not imposed, 
the linear quartic derivative coupling $X$ in $G_4$ 
gives rise to exact hairy BH solutions with an 
asymptotic geometry mimicking the Schwarzschild--(anti-)de Sitter~[(A)dS]
spacetime~\cite{Rinaldi:2012vy,Anabalon:2013oea,Minamitsuji:2013ura,Cisterna:2014nua} (see also Refs.~\cite{Kolyvaris:2011fk,Minamitsuji:2014hha,Cisterna:2015uya,Dong:2017toi,Blazquez-Salcedo:2018tyn,Chatzifotis:2021pak}).
This is an outcome of the violation of the hypothesis~(ii). 
If we consider a quintic-order derivative coupling 
of the form~$G_5\propto \alpha_{\rm GB} \ln |X|$,
which is equivalent to the Gauss-Bonnet 
term~$R_{\rm GB}^2$ linearly coupled to the scalar field~\cite{KYY}, 
there exists an asymptotically flat hairy BH solution
whose metric components are corrected by the 
Gauss-Bonnet coupling~$\alpha_{\rm GB}$~\cite{Sotiriou:2013qea,Sotiriou:2014pfa}. 
This arises from the violation of the hypothesis~(iii).\footnote{ 
We note that there  
exist hairy BH solutions also for 
nonshift-symmetric
Gauss-Bonnet couplings~$\xi (\phi)R_{\rm GB}^2$ 
such as $\xi(\phi) \propto \phi^n$ ($n>1$) and 
$\xi(\phi) \propto {\rm e}^{-\phi}$~\cite{Kanti:1995vq,Kanti:1997br,Doneva:2017bvd,Silva:2017uqg,Antoniou:2017acq,Minamitsuji:2018xde}.}
Another asymptotically flat BH solution violating the hypothesis~(iii) exists in the model where $G_4(X)$ contains $(-X)^{1/2}$~\cite{Babichev:2017guv}.

The linear stability of BHs 
with a time-dependent scalar field 
has been extensively studied in the 
literature~\cite{Ogawa:2015pea,Takahashi:2015pad,Takahashi:2016dnv,Babichev:2017lmw,Tretyakova:2017lyg,Babichev:2018uiw,Takahashi:2019oxz,Khoury:2020aya,Takahashi:2021bml}.
On the other hand, it is yet unclear whether 
the nonasymptotically flat BHs arising from 
the violation of the hypothesis~(ii) or (iii) 
are stable against perturbations 
about the static and spherically symmetric background.
The perturbations of static and spherically symmetric 
BHs in full Horndeski theories in the presence of a time-independent background scalar field
were investigated for both odd-parity~\cite{Kobayashi:2012kh} 
and even-parity~\cite{Kobayashi:2014wsa} sectors.
In these references, the authors obtained conditions 
for the absence of ghosts and Laplacian instabilities 
for high-momentum modes, except the angular 
stability condition of even-parity perturbations. 
Recently, the authors of Ref.~\cite{Kase:2021mix} generalized the results of Refs.~\cite{Kobayashi:2012kh,Kobayashi:2014wsa} by taking into account a perfect fluid, in which the propagation speeds of gravity and scalar-field sectors along the angular directions were also derived.
The linear stability conditions given in Ref.~\cite{Kase:2021mix} 
can be applied not only to BHs but also to neutron stars with 
nontrivial scalar hair~\cite{Kase:2020qvz,Kase:2020yjf}.

In this paper, we study the linear stability of static and 
spherically symmetric BHs in shift-symmetric 
Horndeski theories arising from the violation
of the hypothesis~(ii) or (iii).
We keep the hypothesis~(i), so that 
the background scalar field has a static 
configuration~$\phi(r)$.
We show that the BH solutions 
in reflection-symmetric subclass of shift-symmetric 
Horndeski theories possessing only two coupling functions~$G_2(X)$ and $G_4(X)$ are generically 
prone to the Laplacian instability of even-parity 
perturbations around the horizon. 
In particular, this instability shows up for an exact nonasymptotically flat BH present for theories with 
$G_4 \supset X$~\cite{Rinaldi:2012vy,Anabalon:2013oea,Minamitsuji:2013ura,Cisterna:2014nua} as well as for an asymptotically flat 
BH arising 
in theories with $G_4 \supset (-X)^{1/2}$ \cite{Babichev:2017guv}.\footnote{Here and in what follows, by $G_4\supset (-X)^p$, we mean that the nonminimal derivative coupling~$G_4$ contains a term proportional to $(-X)^p$ on top of the constant term corresponding to the Einstein-Hilbert term.}
We also study several examples of BHs in shift-symmetric Horndeski theories in the presence of the coupling functions~$G_3(X)$ and $G_5(X)$, that break 
the reflection symmetry.
As a first example, we consider a nonasymptotically flat 
BH in GR with a cubic Galileon ($G_3 \propto X$) and 
show that the solution satisfies all the 
stability conditions of linear perturbations. 
Thus, there exists a stable hairy BH 
arising from the violation of the hypothesis~(ii).
As a second example, we study the case with the quintic coupling~$G_5$ containing positive powers of $X$, for which 
it is difficult to realize stable BHs with nontrivial 
scalar hair.
Finally, we investigate the case of 
a scalar field
linearly coupled to 
the Gauss-Bonnet curvature invariant, 
which corresponds to 
$G_5\propto \alpha_{\rm GB} \ln |X|$.
In this case, there is an asymptotically flat 
BH with a finite field derivative $\phi'(r)$ 
on the horizon~\cite{Sotiriou:2013qea,Sotiriou:2014pfa}. 
In the regime of small couplings~$\alpha_{\rm GB}$, 
we show that all the linear stability conditions 
are consistently satisfied for 
this solution.

The rest of this paper is constructed as follows.
In Sec.~\ref{scasec}, we review the shift-symmetric 
Horndeski theories and the properties of static and 
spherically symmetric solutions in vacuum.
In Sec.~\ref{stasec}, we revisit linear 
stability conditions for the odd- and even-parity 
perturbations derived in 
Refs.~\cite{Kobayashi:2012kh,Kobayashi:2014wsa,Kase:2021mix}.
In Sec.~\ref{stealthsec}, we study the stability of GR 
BH solutions with a trivial scalar-field profile~$\phi'(r)=0$.
In Sec.~\ref{refsec}, we show that the hairy BHs with 
nonvanishing scalar-field derivatives appearing in 
the reflection-symmetric 
theories are generically unstable. 
In Sec.~\ref{cuqusec}, we investigate the linear stability 
of hairy BHs arising in nonreflection-symmetric theories.
The last Sec.~\ref{concludesec} is devoted to 
giving a brief summary and conclusion.

%%%%%%%%%%%%%%%%%%%%%%%%%%%%%%%%%%%%%%%%%%
\section{Shift-symmetric Horndeski theories}
\label{scasec}
%%%%%%%%%%%%%%%%%%%%%%%%%%%%%%%%%%%%%%%%%%

The action of shift-symmetric Horndeski theories
 is given by \cite{Horndeski,Def11,KYY,Charmousis:2011bf}
\be
{\cal S}=\int {\rm d}^4 x \sqrt{-g}\,{\cal L}_H\,,
\label{action}
\ee
where $g$ is the determinant of the metric tensor~$g_{\mu \nu}$ and 
\ba
\hspace{-0.5cm}
{\cal L}_H
&=&
G_2(X)-G_{3}(X)\square\phi 
+G_{4}(X)\, R +G_{4,X}(X)\left[ (\square \phi)^{2}
-(\nabla_{\mu}\nabla_{\nu} \phi)
(\nabla^{\mu}\nabla^{\nu} \phi) \right]
\notag\\
&&
+G_{5}(X)G_{\mu \nu} \nabla^{\mu}\nabla^{\nu} \phi
-\frac{1}{6}G_{5,X}(X)
\left[ (\square \phi )^{3}-3(\square \phi)\,
(\nabla_{\mu}\nabla_{\nu} \phi)
(\nabla^{\mu}\nabla^{\nu} \phi)
+2(\nabla^{\mu}\nabla_{\alpha} \phi)
(\nabla^{\alpha}\nabla_{\beta} \phi)
(\nabla^{\beta}\nabla_{\mu} \phi) \right]\,,
\label{LH}
\ea
with $R$ and $G_{\mu \nu}$ being the Ricci scalar 
and Einstein tensor, respectively.
The four functions~$G_{j}$'s ($j=2,3,4,5$) depend only on  
the kinetic term~$X=-g^{\mu\nu}\nabla_{\mu}\phi\nabla_{\nu}\phi/2$, 
with the covariant derivative operator~$\nabla_{\mu}$. 
We also use the notations~$\square \phi \equiv \nabla^{\mu}\nabla_{\mu} \phi$, and 
$G_{j,X} \equiv {\rm d} G_j/{\rm d} X$, 
$G_{j,XX} \equiv {\rm d}^2 G_j/{\rm d} X^2$, etc.

We study static and spherically symmetric solutions in 
shift-symmetric Horndeski theories. 
The metric and scalar field are assumed to be of the following form:
\begin{align}
\rd s^2=-f(r) \rd t^{2} +h^{-1}(r) \rd r^{2}
+ r^{2} \left(\rd \theta^{2}+\sin^{2}\theta\,\rd\varphi^{2} 
\right)\,, \qquad
\phi=\phi(r)\,,
\label{BGmetric}
\end{align}
where $t$, $r$, $(\theta,\varphi)$ are the temporal, radial, 
and angular coordinates, respectively.
The background configuration is characterized by the 
three functions of $r$, i.e., $f(r)$, $h(r)$, and $\phi(r)$.
Note that, on the background~\eqref{BGmetric}, 
the kinetic term of the scalar field can be written as
\be
X=-\frac{1}{2}h\phi'^2\,.
\label{x_def}
\ee
We note that our ansatz~\eqref{BGmetric} corresponds to the hypothesis~\eqref{item1}
mentioned in Sec.~\ref{introsec}.
The independent equations are the $tt$-, $rr$-, and 
$\theta\theta$-components of the equations of 
motion for $g_{\mu\nu}$, which are respectively given by~\cite{Kobayashi:2012kh,Kobayashi:2014wsa,Kase:2021mix}
\ba
{\cal E}_{tt}&\equiv&
\left(A_1+\frac{A_2}{r}+\frac{A_3}{r^2}\right)\phi''
+\left(\frac{\phi'}{2h}A_1+\frac{A_4}{r}+\frac{A_5}{r^2}\right)h'
+G_2-\frac{2 G_{4,X} h^2 \phi'^2+2 G_4 ( h-1)}{r^2}=0
\,,\label{back1}\\
{\cal E}_{rr}&\equiv&
-\left(\frac{\phi'}{2h}A_1+\frac{A_4}{r}+\frac{A_5}{r^2}\right) \frac{hf'}{f}
-\frac{2\phi'}{r}A_1-\frac{1}{r^2}\left[\frac{\phi'}{2h}A_2+(h-1)A_4\right]
-G_2-h G_{2,X} \phi'^2=0
\,,\label{back2}\\
{\cal E}_{\theta\theta}&\equiv&
\left\{\left[A_2+\frac{(2h-1)\phi'A_3+2hA_5}{h\phi' r}\right]\frac{f'}{4f}+A_1+\frac{A_2}{2r}\right\}\phi''
+\frac{1}{4f}\left(2hA_4-\phi'A_2+\frac{2hA_5-\phi'A_3}{r}\right)\left(f''-\frac{f'^2}{2f}\right)\notag\\
&&
+\left[A_4+\frac{2h(2h+1)A_5-\phi'A_3}{2h^2r}\right]\frac{f'h'}{4f}
-\frac{h^2 G_{4,X}\phi'^2+hG_4}{r} \frac{f'}{f}
+\left(\frac{\phi'}{h}A_1+\frac{A_4}{r}\right)\frac{h'}{2}+G_2=0\,,
\label{back3}
\ea
where a prime represents the derivative 
with respect to $r$, and we have defined the 
quantities~$A_1,\cdots,A_5$ 
in Eq.~\eqref{A1_A5}. 
These equations are combined to give
\be
\frac{f'}{2f}{\cal E}_{tt}+{\cal E}_{rr}'
+\left(\frac{f'}{2f}+\frac{2}{r}\right)
{\cal E}_{rr}+\frac{2}{r}{\cal E}_{\theta\theta}
=0\,,
\label{back4}
\ee
which corresponds to the scalar-field equation of motion 
obtained by varying the action~(\ref{action}) with 
respect to $\phi$. 
This is due to the Noether identity associated with 
diffeomorphism invariance~\cite{Motohashi:2016prk}.
More explicitly, Eq.~(\ref{back4}) can be written as
\be
\frac{\rd}{\rd r} \left( r^2 \sqrt{\frac{f}{h}} 
J^r \right)=0\,,
\label{Jreq}
\ee
where $J^r$ is a radial component of 
the Noether current~$J^{\mu}$ associated 
with shift symmetry, given by 
\be
J^r = h \phi' {\cal J}\,, \label{Jr_calJ}
\ee
with 
\ba
{\cal J} & \equiv&
G_{2,X}-\left( \frac{2}{r}+\frac{f'}{2f} 
\right) h \phi' G_{3,X}+2 \left( \frac{1-h}{r^2}
-\frac{h f'}{rf} \right) G_{4,X}+2h \phi'^2 
\left( \frac{h}{r^2}+\frac{hf'}{rf} \right) G_{4,XX} 
\nonumber \\
&&-\frac{f'}{2r^2f} (1-3h)h \phi' G_{5,X}
-\frac{f' h^3 \phi'^3}{2r^2 f}G_{5,XX} \,.
\label{J}
\ea
We note that with the ansatz~\eqref{BGmetric}
$J^r$ is the only nonvanishing component of $J^{\mu}$.
The solution to Eq.~(\ref{Jreq}) is expressed in the form 
\be
J^r(r)=\frac{Q}{r^2} \sqrt{\frac{h}{f}}\,,
\label{Jrso}
\ee
where $Q$ is an integration constant corresponding to 
a scalar charge.
Then, the current strength squared reads
\be
J^2
\equiv
g_{\mu\nu}
J^\mu 
J^\nu
=g_{rr}(J^r)^2 =\frac{Q^2}{r^4f}\,.
\ee
Requiring that $J^2$ is finite 
on the horizon ($f=0$), 
which is the hypothesis~\eqref{item2} mentioned in Sec.~\ref{introsec},
the constant~$Q$ should vanish.
In this case, we have
\be
J^r(r)=h \phi' {\cal J}=0\,,
\label{Jr}
\ee
for any value of $r$. 
We mainly study solutions with $Q=0$ satisfying Eq.~\eqref{Jr}.
Note that, for the Gauss-Bonnet term linearly coupled to a scalar field \cite{Sotiriou:2013qea,Sotiriou:2014pfa}, the divergence of $J^2$ does not necessarily 
invoke unphysical properties of the BH solution~\cite{Creminelli:2020lxn}, and hence a nonvanishing $Q$ is allowed in this particular case (see Sec.~\ref{quinsec}).

Provided that ${\cal J}$ 
is finite in the limit of $\phi'\to 0$,
which is the case when all $G_j$'s are analytic functions of $X$,
namely, all $G_j$'s contain only the zero or positive integer powers of $X$
[the hypothesis~\eqref{item3} in Sec.~\ref{introsec}], 
there are two branches of solutions to Eq.~(\ref{Jr}).
One is the branch~$\phi'=0$, 
while the other is a nontrivial branch satisfying 
\be
{\cal J}=0\,.
\label{Jeq}
\ee
If we impose the asymptotic 
flatness ($f \to 1, h \to 1, f' \to 0, h' \to 0$ as 
$r \to \infty$) with a vanishing field derivative~$\phi'(\infty)=0$
and 
that the contribution of the ordinary kinetic term in $G_2$, 
namely $G_{2,X}(0) X$ with $G_{2,X}(0)\neq 0$, 
is dominant in ${\cal J}$ in the large-$r$ limit,
which is the hypothesis~\eqref{item3} mentioned in Sec.~\ref{introsec},
theories with analytic coupling functions
lead to ${\cal J}\to G_{2,X}(0)$ at spatial infinity. 
Hence, ${\cal J}$ approaches a nonvanishing constant, 
meaning that the branch ${\cal J}=0$ is not present.
In this case, we end up with the no-hair branch with 
$\phi'(r)=0$~\cite{Hui:2012qt}.

On the other hand, if the asymptotic flatness is not imposed 
the derivative~$f'(r)$ can 
be a growing function of $r$. Then, it is possible that terms arising from 
the derivative couplings~$G_3, G_4, G_5$ in Eq.~(\ref{J}) balance the 
term~$G_{2,X}$ to realize ${\cal J}=0$. 
The BH solution present for a quartic-order linear derivative 
coupling $X$ in $G_4$ is such an 
example~\cite{Rinaldi:2012vy,Anabalon:2013oea,Minamitsuji:2013ura,Cisterna:2014nua}. 
The other possibilities for realizing BH solutions 
with $\phi'\neq 0$ are
that $G_2$, $G_3$, $G_4$, and $G_5$ contain $\ln|X|$, 
or fractional/inverse powers of 
$X$~\cite{Sotiriou:2013qea,Sotiriou:2014pfa,Babichev:2017guv},
where the derivatives~$G_{2,X}$, $G_{3,X}$, $G_{4,X}$, $G_{4,XX}$, $G_{5,X}$, and $G_{5,XX}$ have inverse powers of $\phi'(r)$
and their contributions to ${\cal J}$ balance that of the canonical kinetic term
in the limit~$\phi' (r)\to 0$.

%%%%%%%%%%%%%%%%%%%%%%%%%%%%%%%%%%%%%%%%%%
\section{Black hole linear stability conditions}
\label{stasec}
%%%%%%%%%%%%%%%%%%%%%%%%%%%%%%%%%%%%%%%%%%

In order to discuss the linear stability of BHs on the background~(\ref{BGmetric}), 
it is useful to separate perturbations into the odd- and even-parity sectors depending on the 
transformation properties under the rotation in two-dimensional plane~$(\theta, \varphi)$~\cite{Regge:1957td,Zerilli:1970se}. 
In full Horndeski theories with a time independent 
scalar field, the stability conditions against linear perturbations in the odd- and even-parity sectors were derived for BHs~\cite{Kobayashi:2012kh,Kobayashi:2014wsa} 
and relativistic stars~\cite{Kase:2021mix} 
(see also 
Refs.~\cite{DeFelice:2011ka,Motohashi:2011pw,Kase:2014baa,Kase:2020qvz}).
The angular propagation speed of even-parity perturbations 
was not obtained in Ref.~\cite{Kobayashi:2014wsa}, 
but this issue was addressed in Ref.~\cite{Kase:2021mix}.
It should be noted that the linear stability conditions
in Ref.~\cite{Kase:2021mix} 
were derived in the presence of a perfect fluid with density~$\rho$ 
and pressure~$P$ to model static and spherically symmetric stars, and the stability conditions for BHs follow
by taking the limits~$\rho \to 0$ and $P \to 0$.
In the following, 
we summarize the linear stability 
conditions for both odd- and even-parity perturbations.

In the odd-parity sector, the quadratic action for higher multipoles~$\ell\ge 2$ can be written in the form~\cite{Kobayashi:2012kh}
    \be
    {\cal S}_{\rm odd}=\int {\rm d}t{\rm d}r\left[\frac{1}{2}\odd{K}\dot{\chi}^2-\frac{1}{2}\odd{G}\chi'^2-\frac{\ell(\ell+1)}{2}\odd{W}\chi^2-\frac{1}{2}\odd{M}\chi^2\right]\,, \label{qac_odd}
    \ee
where $\chi$ is the master variable and a dot denotes the derivative with respect to $t$, and we have performed the integration over the angular variables.
Here, the coefficients~$\odd{K}$, $\odd{G}$, 
$\odd{W}$, and $\odd{M}$ are determined by the coupling functions in Eq.~\eqref{LH} and the background solution.
Although we do not present the explicit form of 
coefficients, we summarize below the stability conditions which can be read off from the quadratic action.
Note that the boundedness of a Hamiltonian is a coordinate-dependent concept~\cite{Babichev:2018uiw}, and there is a subtlety when the action contains a cross term of time and spatial derivatives~\cite{Takahashi:2019oxz}, which happens for a time-dependent scalar profile~$\phi=qt+\Phi(r)$ with $q \neq 0$. 
In such a case, one should change the coordinate system to remove the cross term. 
For the time-independent scalar field we are 
considering now, the cross term is absent from 
the outset, and hence there is no such an ambiguity.
The ghost-free condition is given by $\odd{K}>0$, 
which reads
\be
{\cal G} \equiv 2 G_4+2 h\phi'^2G_{4,X}
-\frac{f' h^2 \phi'^3 G_{5,X}}{2f}>0\,.
\label{con1}
\ee
For high-momentum modes, the squared propagation speeds 
of odd-parity perturbations along the radial and angular directions are given, respectively, by 
\be
c_{r,{\rm odd}}^2=\frac{g_{rr}}{|g_{tt}|}
\frac{\odd{G}}{\odd{K}}=\frac{{\cal G}}{{\cal F}}\,,
\qquad 
c_{\Omega,{\rm odd}}^2=\frac{g_{\theta\theta}}
{|g_{tt}|}\frac{\odd{W}}{\odd{K}}
=\frac{{\cal G}}{{\cal H}}\,,
\label{csq_odd}
\ee
where 
\ba
{\cal H}&\equiv&2 G_4+2 h\phi'^2G_{4,X}
-\frac{h^2 \phi'^3 G_{5,X}}{r}
\,,\label{calH}\\
{\cal F}&\equiv&2 G_4-h\phi'^2  
\left( \frac12 h' \phi'+h \phi'' \right) G_{5,X}\,.
\label{calF}
\ea
Note that the factor~$\ell(\ell+1)$ in the quadratic action~\eqref{qac_odd} originates from the spherical Laplacian, and hence the coefficient~$\odd{W}$ is associated with the angular propagation speed.
It should also be noted that the squared sound speeds defined in this way are independent of the choice of coordinates.
Under the no-ghost condition~(\ref{con1}), 
the Laplacian instabilities can be avoided for 
\ba
& &
{\cal H}>0\,,
\label{con2}\\
& &
{\cal F}>0\,.
\label{con3}
\ea

In the even-parity sector, the quadratic action for 
higher multipoles~$\ell\ge 2$ can be written in the form~\cite{Kobayashi:2014wsa}
\be
{\cal S}_{\rm even}=
\int {\rm d}t{\rm d}r\sum_{I,J=1}^{2}\left(\frac{1}{2}
{\bm K}_{IJ}\dot{v}^I\dot{v}^J
-\frac{1}{2}
{\bm G}_{IJ}v^{I\prime}v^{J\prime}
-{\bm Q}_{IJ}v^Iv^{J\prime}
-\frac{1}{2}{\bm M}_{IJ}v^Iv^J\right)\,. \label{qac_even}
\ee
Here, $v^I=(\psi,\delta\phi)$ are the master variables, with $\psi$ and $\delta\phi$ corresponding to gravitational field and scalar-field perturbations, respectively.
The coefficient matrices~${\bm K}$, ${\bm G}$, 
${\bm Q}$, and ${\bm M}$ are determined by the coupling functions in Eq.~\eqref{LH} and the background solution, and we choose the overall numerical factor so that the components of ${\bm K}$ are finite in the limit $\ell\to \infty$.
Note that the matrices~${\bm K}$, ${\bm G}$, 
and ${\bm M}$ are symmetric and ${\bm Q}$ is antisymmetric.
Although we do not present the explicit form of coefficient matrices, we summarize below the stability conditions which can be read off from the quadratic action.
Ghost instabilities can be avoided if both the eigenvalues of ${\bm K}$ are positive, i.e.,
\be
{\bm K}_{11}>0 \quad {\rm and} \quad
\det {\bm K}>0\,.
\ee
Provided that the condition~\eqref{con3} holds, these conditions can be satisfied if
\be
{\cal K} \equiv 2{\cal P}_1-{\cal F}>0\,,
\label{con4}
\ee
where 
\be
\label{p1mu}
{\cal P}_1 = \frac{h \mu}{2fr^2 {\cal H}^2} \left( 
\frac{fr^4 {\cal H}^4}{\mu^2 h} \right)'\,,\qquad 
\mu = \frac{2(\phi' a_1+r \sqrt{fh}\,{\cal H})}{\sqrt{fh}}\,,
\ee
and $a_1$ is defined in 
Eq.~\eqref{a1}.
In the limit of high frequencies, the squared radial propagation speeds of $\psi$ and $\delta\phi$ are given 
as eigenvalues of the matrix
\be
{\bm c}_{r,{\rm even}}^2\equiv 
\frac{g_{rr}}{|g_{tt}|}
{\bm K}^{-1}{\bm G}\,.
\ee
Written explicitly, 
we have~\cite{Kobayashi:2012kh,Kobayashi:2014wsa,Kase:2021mix}
\ba
c_{r1,{\rm even}}^2 &=&
\frac{\cal G}{\cal F}\,,
\label{cr1}\\
c_{r2,{\rm even}}^2 &=&
\frac{2\phi'[ 4r^2 (fh)^{3/2} {\cal H} c_4 
(2\phi' a_1+r\sqrt{fh}\,{\cal H})
-2a_1^2 f^{3/2} \sqrt{h} \phi' {\cal G} 
+( a_1 f'+2 c_2 f ) r^2 fh 
{\cal H}^2]}{f^{5/2} h^{3/2} 
(2{\cal P}_1-{\cal F}) \mu^2}
\,,\label{cr2}
\ea
where $c_2$ and $c_4$ are defined in Eqs.~\eqref{c2} and \eqref{c4}.
Since $c_{r1,{\rm even}}^2$ is identical to 
$c_{r,{\rm odd}}^2$ in Eq.~\eqref{csq_odd}, the conditions~(\ref{con1}) 
and (\ref{con3}) ensure that $c_{r1,{\rm even}}^2>0$.
In order to avoid the Laplacian instability of $\delta \phi$ 
along the radial direction, we require that 
\be
c_{r2,{\rm even}}^2>0\,.
\label{con5}
\ee
For the monopole mode ($\ell=0$), there is no propagation 
for the gravitational perturbation~$\psi$, while the scalar-field perturbation~$\delta \phi$ 
propagates with the same radial 
velocity as Eq.~(\ref{cr2}). 
For the dipole mode ($\ell=1$), there is 
a gauge degree 
of freedom for fixing $\delta \phi=0$, 
under which the perturbation $\psi$ propagates 
with the same radial speed squared as Eq.~(\ref{cr2}). 
The squared angular propagation speeds of $\psi$ and $\delta \phi$ in the large-$\ell$ limit are obtained as eigenvalues of the matrix
\be
    {\bm c}_{\Omega,{\rm even}}^2\equiv \lim_{\ell\to\infty}\frac{1}{\ell(\ell+1)}
    \frac{g_{\theta\theta}}{|g_{tt}|}{\bm K}^{-1}
    {\bm M}\,.
\ee
This gives the following biquadratic equation~\cite{Kase:2021mix}:
\be
c_{\Omega}^4+2B_1c_{\Omega}^2+B_2=0\,,
\ee
namely,
\be
c_{\Omega \pm}^2=-B_1\pm\sqrt{B_1^2-B_2}\,,
\label{cosq}
\ee
where $B_1$ and $B_2$ are defined in Eqs.~\eqref{B1def} and \eqref{B2def}.
The branch of the square root in Eq.~\eqref{cosq} is chosen so that 
$c_{\Omega \pm}^2$ are smooth functions 
of $r$ [see also the comment below Eq.~\eqref{csq1_cubic}].
Note that the above squared sound speeds of even-parity perturbations along the radial and angular directions are scalar quantities independent of the choices of gauges and coordinates.
The Laplacian instabilities along the angular 
direction can be avoided for 
\ba
& &
c_{\Omega +}^2>0\,,
\label{con7}\\
& &
c_{\Omega -}^2>0\,.
\label{con8}
\ea
From Eq.~\eqref{cosq},
these conditions are realized if
\be
B_1^2\ge B_2>0 \quad
\text{and} \quad
B_1<0\,.
\label{B1B2con}
\ee

In summary, there are neither ghosts nor 
Laplacian instabilities under the 
conditions~(\ref{con1}), (\ref{con2}), (\ref{con3}), 
(\ref{con4}), (\ref{con5}), 
and \eqref{B1B2con}.
In \hyperref[table]{Table}, we summarize these 
stability conditions for convenience.

\vskip-\baselineskip
\begin{table}[ht]
\renewcommand\thetable{\!\!}
\newcommand\xrowht[2][0]{\addstackgap[.5\dimexpr#2\relax]{\vphantom{#1}}}
\newcolumntype{C}[1]{>{\hfil}m{#1}<{\hfil}}
\centering
\caption{Summary of the linear stability conditions.}
\begin{tabular}{C{30mm}C{25mm}C{25mm}C{45mm}} \hline\hline \xrowht{12pt}
     & No ghost & $c_r^2>0$ & $c_\Omega^2>0$ \\ \hline \xrowht{12pt}
    Odd modes & ${\cal G}>0$ & ${\cal F}>0$ & ${\cal H}>0$ \\ \xrowht{12pt}
    Even modes & ${\cal K}>0$ & $c_{r2,{\rm even}}^2>0$ & $B_1^2\ge B_2>0$ and $B_1<0$ \\ \hline\hline
\end{tabular}\label{table}
\end{table}

%%%%%%%%%%%%%%%%%%%%%%%%%%%%%%%%%%%%%%%%%%
\section{Linear stability of general relativity solutions}
\label{stealthsec}
%%%%%%%%%%%%%%%%%%%%%%%%%%%%%%%%%%%%%%%%%%

First of all, we study the linear stability of 
BH solutions with a trivial scalar-field 
profile, i.e., 
\be
\phi'=0\,.
\label{phiva}
\ee
As mentioned in Sec.~\ref{scasec}, such a solution exists 
as long as ${\cal J}$ is finite
in the limit that $\phi'\to 0$.
We shall discuss the other branch of hairy BH solution 
satisfying ${\cal J}=0$ in the next sections.
Note that the functions~$G_j$'s and their derivatives are evaluated at $X=0$ throughout this section.
In this case, the background equations~\eqref{back1}--\eqref{back3} 
reduce to the following 
two independent equations:
\begin{align}
hf'-fh'&=0\,, \\
2G_4 h \frac{(rf)'}{r^2f}-G_2-\frac{2G_4}{r^2}&=0\,.
\end{align}
They can be solved to yield
\be
h=C_0f=1-\frac{r_0}{r}+\frac{G_2}{6G_4}r^2\,,
\label{Schde}
\ee
with $C_0$ and $r_0$ being integration constants.
The integration constant~$C_0$ can be absorbed into a rescaling of $t$, and hence we obtain the Schwarzschild-(A)dS metric, i.e., the BH solution 
in GR.
Nevertheless, due to the existence of a dynamical scalar field, the stability under linear perturbations is rather nontrivial, as we shall see below.

Let us first discuss the linear stability of no-hair 
BHs against odd-parity perturbations.
For $\phi'=0$, the quantities relevant to 
stability of odd modes are simply given by 
\be
{\cal F}={\cal G}={\cal H}=2G_4\,.
\ee
Therefore, the no-ghost condition~\eqref{con1} yields
\be
G_4>0\,. 
\label{con1_stealth}
\ee
Also, we have
\be
c_{r,{\rm odd}}^2=c_{\Omega,{\rm odd}}^2=1\,,
\ee
meaning that there is no 
Laplacian instability.

Next, we study the BH stability against 
even-parity perturbations.
A caveat here is that the quantities 
associated 
with stability of even modes contain $\phi'$ in their denominators, though one can obtain finite $\phi'\to 0$ limits.
In order to remove vanishing $\phi'$ in the denominator, one should redo the computation of Refs.~\cite{Kobayashi:2014wsa,Kase:2021mix} 
with $\phi'$ set to zero from the outset, 
but this reproduces 
the $\phi'\to 0$ limits of the final results.
Hence, one can safely take the $\phi'\to 0$ limit.
Another point to note is that the 
quantity~${\cal K}$, 
which is associated with the no-ghost condition of 
even-parity perturbations, is vanishing in the limit~$\phi'\to 0$.
This could be a problem because it may imply the degeneracy of the kinetic matrix, which leads to strong coupling.
However, as pointed out in Ref.~\cite{Kobayashi:2014wsa}, 
the determinant of the kinetic matrix is proportional to ${\cal K}/\phi'^2$, which is finite in the limit~$\phi'\to 0$, 
meaning that the strong coupling problem is actually absent. 
To be more concrete, we solve Eqs.~(\ref{back1})--(\ref{back3}) 
for $h'$, $f'$, $f''$, substitute them into ${\cal K}$, and 
finally take the limits $\phi'\to 0$ and $\phi'' \to 0$.
Then, the no-ghost condition is read off from
\be
\lim_{\phi'\to 0}\frac{\cal K}{\phi'^2}
=\frac{G_{2,X} G_4-G_2 G_{4,X}}{2G_4}r^2>0\,.
\ee
On using the inequality (\ref{con1_stealth}), 
this condition translates to
\be
G_{2,X} G_4-G_2 G_{4,X}>0\,. 
\label{con2_stealth}
\ee
If we consider the Schwarzschild BH solution, 
we have $G_2=0$ in Eq.~(\ref{Schde}). 
In this case,  
the condition (\ref{con2_stealth}) reduces to 
$G_{2,X}G_4>0$, which is consistent with the 
one derived in Ref.~\cite{Kobayashi:2014wsa}.
The squared sound speeds in the radial direction reduce to
\be
c_{r1,{\rm even}}^2=c_{r2,{\rm even}}^2=1\,.
\ee
Moreover, we have $B_1=-1$ and $B_2=1$, 
so that 
\be
c_{\Omega+}^2=c_{\Omega-}^2=1\,.
\ee

In summary, the Schwarzschild-AdS solution is free of ghosts/Laplacian instabilities if the conditions~\eqref{con1_stealth} and \eqref{con2_stealth} are satisfied, with the propagation speeds equivalent to that of light.

%%%%%%%%%%%%%%%%%%%%%%%%%%%%%%%%%%%%%%%%%%
\section{Generic instability for reflection-symmetric theories}
\label{refsec}
%%%%%%%%%%%%%%%%%%%%%%%%%%%%%%%%%%%%%%%%%%

We investigate the linear stability of BH solutions in shift- and reflection-symmetric Horndeski theories, for which $G_3=G_5=0$.
Namely, we consider shift-symmetric Horndeski theories containing two arbitrary functions~$G_2(X)$ and $G_4(X)$ with the Lagrangian 
\be
{\cal L}_{H}=G_2(X) 
+G_{4}(X)\, R +G_{4,X}(X)\left[ (\square \phi)^{2}
-(\nabla_{\mu}\nabla_{\nu} \phi)
(\nabla^{\mu}\nabla^{\nu} \phi) \right]\,.
\ee
We focus on BH solutions with a nontrivial scalar-field profile $\phi'\ne 0$ satisfying Eq.~\eqref{Jeq}, so that 
\be
{\cal J}=G_{2,X}+2 \left( \frac{1-h}{r^2}
-\frac{h f'}{rf} \right) G_{4,X}+2h \phi'^2 
\left( \frac{h}{r^2}+\frac{hf'}{rf} \right) 
G_{4,XX}=0\,. 
\label{calJ2}
\ee
An explicit example of BH solutions of 
this type is present for 
$G_2(X)=\eta X-\Lambda$ and 
$G_4(X)=\Mpl^2/2-\alpha_1 X/2$ \cite{Rinaldi:2012vy,Anabalon:2013oea,Minamitsuji:2013ura}. 
As long as $G_{2,X}$, $G_{4,X}$, and $G_{4,XX}$ 
do not contain fractional or negative powers 
of $\phi'$, there is the trivial GR solution 
$\phi'=0$ besides the branch (\ref{calJ2}). 
The linear stability of BH solutions for the 
GR branch was already studied 
in Sec.~\ref{stealthsec}.

For the branch satisfying Eq.~(\ref{calJ2}), the background equations of motion 
can be simply expressed as~\cite{Kobayashi:2014eva}
\begin{align}
8X \left(G_{4,X}^2+G_{4}G_{4,XX} \right)
&=r^2 \left({\cal G}G_2 \right)_{,X}\,, 
\label{X_reflection} \\
\left( r{\cal G}^2h \right)'&
={\cal G} \left( 2G_4+r^2 G_2 \right)\,, 
\label{h_reflection} \\
\left(\frac{f}{{\cal G}^2h}\right)'&=0\,.
\label{f_reflection} 
\end{align}
We recall that ${\cal G}$ is given by Eq.~\eqref{con1}, 
which in the present case reads
\be
{\cal G}=2 \left( G_{4}-2X G_{4,X} 
\right)>0\,.
\ee
We can solve Eq.~(\ref{X_reflection}) to yield $X$ algebraically 
as a function of $r$.
Then, from Eqs.~(\ref{h_reflection}) and (\ref{f_reflection}), 
we obtain $h$ and $f$ as functions of $r$.
More concretely,
\begin{align}
f&=C_1 {\cal G}^2 h\,, \label{f_reflection2} \\
h&=\frac{1}{r{\cal G}^2} \int_{r_s}^r\, 
\rd r\,{\cal G} \left( 2G_4+r^2 G_2 \right)\,, \label{h_reflection2}
\end{align}
where $C_1\,(\ne 0)$ and $r_s\,(>0)$ are integration constants.
Provided that the integrand in Eq.~\eqref{h_reflection2} approaches 
to a constant as $r \to r_s$, 
we have the following expansions:
\begin{align}
f&=f_1(r-r_s)+f_2(r-r_s)^2
+\cdots\,, \label{fexpan} \\
h&=h_1(r-r_s)+h_2(r-r_s)^2
+\cdots\,, \label{hexpan}
\end{align}
where $f_j$ and $h_j$ ($j=1,2,\cdots$) are constants. 
For $f_1>0$ and $h_1>0$, the coordinate distance~$r_s$ can be 
identified as the position of BH horizon.
If there exists some finite coordinate distance~$\rc\,(>\rBH)$ such that $f>0$ and $h>0$ for $\rBH<r<\rc$ and $f(\rc)=h(\rc)=0$, the radius~$\rc$ can be identified as a cosmological horizon. 
The expansion of $X$ around the BH horizon is given by 
\be
X=X_s+X'(r_s) (r-r_s)
+{\cal O}((r-r_s)^2)\,, \label{Xexpan}
\ee
where $X'(r_s)$ can be evaluated by 
taking the 
$r$-derivative of Eq.~(\ref{calJ2}). 
Note that the value of $X_s$ is obtained algebraically from Eq.~\eqref{X_reflection} with $r=r_s$.
Unless $G_2$ and $G_4$ are fine-tuned, we have $X_s\ne 0$.
The functions~$G_2(X)$, $G_4(X)$, and their $X$-derivatives appearing in 
quantities relevant to the BH stability conditions are also expanded 
around the value of $X=X_s$ at the BH horizon, e.g.,
\be
G_{4}(X)=G_{4}(X_s)+G_{4,X}(X_s)(X-X_s)+{\cal O}((X-X_s)^2)\,. 
\ee

In order to avoid ghosts/Laplacian instabilities, we require that all the linear conditions listed
in \hyperref[table]{Table} are satisfied from the BH horizon to spatial infinity (or the cosmological horizon, 
if it exists).
In the present case, however, it is 
impossible to satisfy all these conditions. 
Taking the product ${\cal F}{\cal K}B_2$ 
in the vicinity of $r=r_s$, we obtain
\be
{\cal F}{\cal K}B_2
=-\frac{4X_s^4(G_{4,X}^2+G_{4}G_{4,XX})^2}{(G_{4}-4X_s G_{4,X}-4X_s^2G_{4,XX})^2}\frac{r_s^2}{(r-r_s)^2}
+{\cal O}((r-r_s)^{-1})\,, 
\label{FKB2_general}
\ee
where $G_4$ and its $X$-derivatives on 
the right-hand side are evaluated at $X=X_s$. Provided that 
\be
X_s \ne 0 \quad {\rm and} \quad 
G_{4,X}^2+G_{4}G_{4,XX}\ne 0\,,
\label{Xhcon}
\ee
the leading-order contribution to 
Eq.~(\ref{FKB2_general}) is negative, i.e., 
\be
{\cal F}{\cal K}B_2<0\,,\quad
{\rm for}~~r \to r_s\,.
\label{FKB2_general2}
\ee
This shows that the quantities~${\cal F}$, ${\cal K}$, and $B_2$ cannot be positive at the same time near the 
BH horizon in general, leading to instability.
For instance, even if the two conditions ${\cal F}>0$ and ${\cal K}>0$ 
are satisfied, we have $B_2<0$, and hence 
\be
c_{\Omega -}^2=-B_1-\sqrt{B_1^2-B_2}<0\,.
\ee
Thus, the angular Laplacian instability of 
even-parity perturbations 
is unavoidable around the BH horizon.
We stress that the knowledge of the angular propagation speeds is essential to recognize the instability of this kind.
A similar instability was found for stealth Schwarzschild-dS solutions with linearly time-dependent scalar hair in degenerate higher-order scalar-tensor theories~\cite{Takahashi:2021bml}.

We note that, from Eq.~\eqref{x_def},
$X<0$ outside the BH horizon [$h(r)>0$] and $X>0$ inside the BH horizon [$h(r)<0$], and hence the character of the scalar field is spacelike outside the horizon and timelike inside the horizon, respectively.
If a BH solution with $X_s<0$ could be extended to the interior of the BH horizon, the coordinate invariant~$X$ would have a sudden change of the sign across the horizon, indicating that the horizon would become a singular hypersurface. Thus, a BH solution with $X_s<0$ cannot be extended to the interior of the BH horizon and can be defined only in the domain outside the horizon where $h(r)>0$ and the character of the scalar 
field is spacelike.
Our result~\eqref{FKB2_general2} suggests that
BH solutions with $X_s\neq 0$ generically suffer from 
instabilities in the domain
where the solution can exist, and hence 
such solutions could not be realistic.
In other words, only the physically acceptable BH 
solution defined in both the exterior and interior of 
the horizon should have $X_s=0$.
We note that a similar argument could be applied 
to the cosmological horizon (if it exists), and
a static and spherically symmetric solution with $X(r_c)<0$
could not be extended to the exterior of the cosmological horizon.

The above instability is generic for the theories satisfying 
the condition~(\ref{Xhcon}). 
In Secs.~\ref{quarposec}--\ref{nhalfsec}, we apply the above results to theories containing positive power-law 
functions~$(-X)^p$ with $p>0$ in $G_4(X)$
and $G_2(X)=\eta X-\Lambda$, with $\eta$ and $\Lambda$ being constant.
Note that the above discussion does not apply if the conditions 
in Eq.~\eqref{Xhcon} are not satisfied, in which case a further analysis is required. Such a situation occurs when, e.g., the nonminimal derivative coupling to the Ricci scalar is absent ($G_4={\rm constant}$).
In this case, as we shall see in Sec.~\ref{k-essence}, 
the perturbations would be strongly coupled.

\subsection{\texorpdfstring{$G_4\supset (-X)^p$ with $p>1$}{X to p in G4}}
\label{quarposec}

As a demonstration of the generic instability, let us study theories given by the coupling functions
\be
G_2=\eta X-\Lambda\,, \qquad
G_4=\frac{\Mpl^2}{2}
+\frac{\alpha_p}{2}(-X)^p\,,
\label{G4power}
\ee
where $\Mpl$ is the reduced Planck mass, and $\eta$, $\Lambda$, $\alpha_p$, and 
$p$ are constants.
Note that we put a minus sign in $(-X)^p$ because $X=-h {\phi'}^2/2$ is negative 
for $h>0$.

For the branch satisfying Eq.~(\ref{calJ2}), 
the kinetic term of the scalar field is expressed as
\be
(-X)^{p-1}=-\frac{\eta r^2 f}
{p \alpha_p [(2p-1)(r f'+f)h-f]}\,.
\label{Xde}
\ee
The linear derivative coupling ($p=1$) 
is a special case in which the left-hand side of Eq.~(\ref{Xde}) is constant. 
In this section, we study the 
power-law models with
\be
p > 1\,,
\ee
which accommodate the quartic Galileons ($p=2$). 
We shall discuss the $p=1$ and $p=1/2$ cases
separately in Secs.~\ref{linearsec} and \ref{nhalfsec}, respectively.

Around the BH horizon~$r=r_s$, the leading-order 
contributions to $f$ and $h$ are 
$f_1 (r-r_s)$ 
and $h_1 (r-r_s)$, respectively. 
This mean that, as $r \to r_s$, the kinetic 
term $X$ approaches 
a constant~$X_s$, satisfying 
\be
(-X_s)^{p-1}=-\frac{\eta r_s^2}
{p\alpha_p [(2p-1)h_1 r_s-1]}\,,
\ee
and hence $X_s \neq 0$. 
As a result, $\phi'^2$ diverges 
on the BH horizon. Due to the property 
$X_s \neq 0$, the term
$G_{4,X}^2+G_{4}G_{4,XX}$ 
appearing in the numerator of 
Eq.~(\ref{FKB2_general}) does not 
generally vanish.
Unless the coupling~$\alpha_p$ is fine-tuned to satisfy $G_{4,X}^2+G_{4}G_{4,XX}\propto (2p-1)\alpha_p (-X_s)^p+(p-1)\Mpl^2=0$,
the product ${\cal F}{\cal K}B_2$ 
around $r=r_s$ yields
\be
{\cal F}{\cal K}B_2=
-\frac{p^2 \alpha_p^2 (-X_s)^{2p}
[(2p-1)\alpha_p (-X_s)^p
+(p-1)\Mpl^2]^2}
{[(4p^2-1)\alpha_p (-X_s)^{p}
-\Mpl^2]^2}
\frac{r_s^2}{(r-r_s)^2}+{\cal O}((r-r_s)^{-1})\,.
\label{pro2}
\ee
The leading-order contribution to Eq.~(\ref{pro2}) is 
negative outside the BH horizon, and hence 
the corresponding BH solutions are unstable in general.

\subsection{\texorpdfstring{$G_4\supset X$}{X in G4}}
\label{linearsec}

Having discussed general power-law quartic derivative couplings 
$G_4\supset (-X)^p$
with $p>1$, we now study the special case~$p=1$, 
for which the coupling functions read
\be
G_2=\eta X-\Lambda\,, \qquad
G_4=\frac{\Mpl^2}{2}-\frac{\alpha_1}{2}X\,.
\label{G4power1}
\ee
In this case, the field 
equations~\eqref{X_reflection}--\eqref{f_reflection} yield 
the following exact 
solution~\cite{Rinaldi:2012vy,Anabalon:2013oea,Minamitsuji:2013ura,Cisterna:2014nua}:
\begin{align}
f&=-\frac{r_0}{r}+\frac{\sqrt{-\alpha_1\eta}
(\Mpl^2\eta-\alpha_1\Lambda)^2}
{4\Mpl^4\eta^3r}{\rm arctan}
\left(-\frac{\sqrt{-\alpha_1 \eta}}{\alpha_1}r\right)
-\frac{(\Mpl^2\eta+\alpha_1 \Lambda)^2}
{12\Mpl^4\alpha_1\eta}r^2
+\frac{(\Mpl^2\eta+\alpha_1 \Lambda)
(3\Mpl^2\eta-\alpha_1\Lambda)}
{4\Mpl^4\eta^2}\,, \nonumber \\
h&=\frac{4\Mpl^4(\eta r^2-\alpha_1)^2}
{[(\Mpl^2\eta+\alpha_1\Lambda)r^2
-2\Mpl^2\alpha_1]^2}f\,, \qquad
X=\frac{(\Mpl^2\eta-\alpha_1\Lambda)r^2}
{2\alpha_1(\alpha_1-\eta r^2)}\,,
\label{Rinaldi_sol}
\end{align}
where we have chosen the integration constants~$\rBH$ and $C_1$ 
in Eqs.~(\ref{h_reflection}) and (\ref{f_reflection}) 
so that $f\simeq 1-r_0/r$ for small $r$.
For the existence of this solution, 
we require
\be
\alpha_1 \eta<0\,.
\ee
Provided that $\Mpl^2\eta-\alpha_1\Lambda \neq 0$, 
there exists a nontrivial branch 
with $X \neq 0$. 
Also, since the character of the scalar 
field is spacelike and $X$
should be negative outside the BH horizon, we have
\be
\Mpl^2\eta-\alpha_1\Lambda<0\,.
\label{licon}
\ee
Let us denote by $X_s\,(<0)$ the value of $X$ 
at the BH horizon~$r=r_s$.
As noted previously,
the solution~\eqref{Rinaldi_sol} could exist only outside the BH horizon,
as the scalar field becomes imaginary inside the horizon.
On the contrary,
if a solution similar to Eq.~\eqref{Rinaldi_sol} exists inside the BH horizon where $h(r)<0$
and the scalar field is timelike (i.e., $X>0$),
it could not be extended to the exterior of the BH horizon.
Then, the  
product~${\cal F}{\cal K}B_2$ yields 
\be
{\cal F}{\cal K}B_2=
-\frac{\alpha_1^4 
X_s^4}{(\Mpl^2+3\alpha_1 X_s)^2}
\frac{r_s^2}{(r-r_s)^2}
+{\cal O} ((r-r_s)^{-1})\,,
\ee
whose leading-order term is negative.
Thus, the exact BH 
solution (\ref{Rinaldi_sol}) 
is excluded by the instability 
problem around the BH horizon.
The instability of the 
solution \eqref{Rinaldi_sol}
is one of our main results.
We note that, in generalized Proca theories~\cite{Chagoya:2016aar,Minamitsuji:2016ydr,Heisenberg:2014rta}
with a vector field~$A_{\mu}$, 
there is an exact Schwarzschild 
solution with a nonvanishing longitudinal vector component 
in the presence of a quartic coupling~$G_4(Y)$ containing a linear function 
of $Y=-A^{\mu}A_{\mu}/2$~\cite{Chagoya:2016aar,Heisenberg:2017xda,Heisenberg:2017hwb}. 
Such BH solutions are also 
prone to a similar instability problem 
of vector-field perturbations in the 
odd-parity sector around 
the horizon \cite{Kase:2018voo}.

For the theory~(\ref{G4power1}), there exists 
a GR branch of the vanishing 
field profile ($\phi'=0$). This branch 
is free from the ghost instability under the condition~(\ref{con2_stealth}), 
i.e., $\Mpl^2\eta-\alpha_1\Lambda>0$, 
which is an opposite inequality to Eq.~(\ref{licon}).
Thus, under the inequality 
$\Mpl^2\eta-\alpha_1\Lambda<0$, 
neither the branch 
$\phi' \neq 0$ nor 
the other branch $\phi'=0$ is stable.

\subsection{\texorpdfstring{$G_4\supset (-X)^{1/2}$}{X to 1/2 in G4}}
\label{nhalfsec}

Let us study another special case with $p=1/2$, i.e.,
\be
G_2=\eta X-\Lambda\,, \qquad
G_4=\frac{\Mpl^2}{2}
+\frac{\alpha_{1/2}}{2}(-X)^{1/2}\,.
\label{G4powerhalf}
\ee
Since 
${\cal J}=(2\eta r^2 X+\alpha_{1/2}
\sqrt{-X})/(2 r^2 X)=0$, the
kinetic term of the scalar field 
corresponding to the branch $\phi' \neq 0$ 
is given by 
\be
X=-\frac{\alpha_{1/2}^2}{4\eta^2 r^4}\,,
\label{Xhalf}
\ee
where we have assumed $\eta \alpha_{1/2}>0$.
A positive power of $\phi'$ is present 
in the denominator of ${\cal J}$, so 
the radial current equation, 
$J^r=h \phi' {\cal J}=0$, does not 
allow the existence of a branch of 
vanishing field derivative ($\phi'=0$).

Integrating Eqs.~(\ref{h_reflection}) and 
(\ref{f_reflection}) with Eq.~(\ref{Xhalf}),  
we obtain the following exact solution~\cite{Babichev:2017guv}
\begin{align}
f=h=1-\frac{r_0}{r}
-\frac{\alpha_{1/2}^2}
{4\Mpl^2 \eta r^2}
-\frac{\Lambda}{3\Mpl^2}r^2\,,
 \label{fhhalf}
\end{align}
where the integration constants have been 
chosen to have the behavior 
$f=h \simeq 1-r_0/r$ for small $r$. 
We note that, 
because of the dependence~$(-X)^{1/2}$ in $G_4$,
the solution~\eqref{fhhalf}
is defined only in the domain where
the scalar field is spacelike (i.e., $X<0$),
which corresponds to the domain 
outside the BH horizon and inside the cosmological horizon (for $\Lambda>0$).
In the case of $\Lambda=0$, 
the solution (\ref{fhhalf}) 
becomes
an asymptotically flat spacetime, 
i.e., $f=h \to 1$ as $r \to \infty$.
On using Eq.~(\ref{Xhalf}), in the vicinity of the BH horizon~$r=r_s$, 
the product~${\cal F}{\cal K}B_2$ reduces to 
\be
{\cal F}{\cal K}B_2
=-\frac{\alpha_{1/2}^4}{64 \eta^2 r_s^4}
\frac{r_s^2}{(r-r_s)^2}
+{\cal O}((r-r_s)^{-1})\,.
\label{FKB2_half}
\ee
Since the leading-order contribution to 
${\cal F}{\cal K}B_2$ is negative for $\alpha_{1/2} \neq 0$, 
the exact solution~(\ref{fhhalf}) with (\ref{Xhalf}) 
inevitably suffers from the 
instability around the BH horizon.

\subsection{Strong coupling for k-essence}
\label{k-essence}

In this subsection, we study the case of k-essence given by the action
\be
{\cal S}=\int {\rm d}^4x\sqrt{-g}\left[G_2(X)+\frac{\Mpl^2}{2}R\right]\,.
\ee
From Eq.~(\ref{calJ2}), the branch of a nonvanishing 
field derivative obeys
\be
G_{2,X}=0\,,
\label{G2X}
\ee
implying that $X$ is constant everywhere.
Also, the value of $X$ is determined as a solution to the (algebraic) equation~\eqref{G2X}.
From the background equations~(\ref{back1}) and 
(\ref{back2}), we obtain
\be
G_2=\frac{(rhf'+fh-f)\Mpl^2}{r^2 f}\,,\qquad
hf'= fh'\,.
\label{backape}
\ee
The quantities associated with the linear stability of BHs yield 
\be
{\cal K}=0\,,\qquad
c_{r2,{\rm even}}^2=\infty\,,
\ee
with ${\cal F}={\cal G}={\cal H}=\Mpl^2$ and $c_{\Omega \pm}^2=1$. 
Such a diverging sound speed typically implies an infinitely strong coupling, and hence the perturbative treatment would no longer be viable in this case.

%%%%%%%%%%%%%%%%%%%%%%%%%%%%%%%%%%%%%%%%%%
\section{Nonreflection-symmetric theories: Case studies}
\label{cuqusec}
%%%%%%%%%%%%%%%%%%%%%%%%%%%%%%%%%%%%%%%%%%

In the previous section, we showed that BH solutions with nontrivial scalar hair in the reflection-symmetric subclass of shift-symmetric Horndeski theories are linearly unstable in general.
In this section, we study the linear stability for several examples of BH solutions 
in nonreflection-symmetric theories
containing the coupling functions~$G_3(X)$ or $G_5(X)$ 
in the Lagrangian. 
In this case, the scalar-field profile around the BH horizon is quite different from the one in the reflection-symmetric case.
Indeed, substituting the expansions~\eqref{fexpan}--\eqref{Xexpan} around the BH horizon into the background equations, one finds that the left-hand side of Eq.~\eqref{back2} behaves as
\be
\sqrt{2h_1}(-X_s)^{3/2}\left[G_{3,X}(X_s)+\frac{1}{r_s^2}G_{5,X}(X_s)\right](r-r_s)^{-1/2}+{\cal O}((r-r_s)^0)\,,
\ee
which in general does not vanish unless
\be
X_s=0 \quad {\rm or} \quad 
G_{3,X}(X_s)+\frac{1}{r_s^2}G_{5,X}(X_s)=0\,.
\label{back2_BHhorizon}
\ee
Therefore, unless $G_3$ and $G_5$ are fine-tuned to satisfy the latter of Eq.~\eqref{back2_BHhorizon}, the kinetic term of the scalar field is vanishing on the BH horizon.
In what follows, we study BHs having $X=0$ on the horizon in the presence of the coupling functions~$G_3$ and $G_5$. 
In Sec.~\ref{cubicsec}, we first consider cubic-order power-law 
couplings~$G_3(X)\propto (-X)^p$
with $p>0$ and focus on the case of cubic Galileons ($p=1$).
In Sec.~\ref{quinsec}, we discuss quintic-order power-law 
couplings~$G_5(X)\propto (-X)^p$
with $p>0$. 
In Sec.~\ref{Gausssec}, we investigate 
the case with the scalar field linearly coupled to the Gauss-Bonnet curvature invariant, 
which amounts to $G_5(X)\propto \ln |X|$.

\subsection{Cubic Galileons}
\label{cubicsec}

We consider theories characterized by the coupling functions
\be
G_2=\eta X-\Lambda\,,\qquad 
G_3=\gamma_p 
\left( -X \right)^p \,,\qquad 
G_4=\frac{\Mpl^2}{2}\,,\qquad 
G_5=0\,,
\ee
where $\eta$, $\Lambda$, $\gamma_p$, 
and $p\,(> 0)$ are constants.
The branch with $\phi'(r) \neq 0$ 
obeys Eq.~(\ref{Jeq}), i.e., 
\be
\eta
+p \gamma_p
\left( \frac{2}{r}+\frac{f'}{2f} \right) h \phi' 
\left(-X \right)^{p-1}=0\,,
\label{G3sca}
\ee
where $X=-h \phi'^2/2$. 
Around the BH horizon~$r=r_s$, 
we expand the metric components~$f$ and $h$ as in Eqs.~(\ref{fexpan}) and (\ref{hexpan}). 
At leading order, the scalar-field derivative~$\phi'$ 
around $r=r_s$ is expressed as
\be
{\phi'}^{2p-1}=-\frac{\eta}{p \gamma_p} 
\left(\frac{h_1}{2}\right)^{-p} 
(r-r_s)^{-(p-1)}\,, 
\label{phide}
\ee
and hence
\be
X\propto (r-r_s)^{1/(2p-1)}\,.
\ee
For $p>1/2$, the kinetic term~$X$ vanishes 
on the horizon. 
This property is different from that in 
the case of $G_4(X)\supset (-X)^p$ studied in Sec.~\ref{quarposec}, for which $X(r_s)<0$.

From Eq.~(\ref{phide}), we find that 
the cubic Galileons ($p=1$) correspond to a special case 
in which $\phi'$ approaches a nonvanishing constant as $r \to r_s$.
For $p>1$, the scalar-field derivative diverges as 
$\phi' \propto (r-r_s)^{-(p-1)/(2p-1)}$ 
on the horizon.
In the following, we study the linear stability of BHs 
in cubic Galileons, 
i.e., 
\be
\label{quar2}
G_2=\eta X\,,\qquad 
G_3=\gamma_1(- X) \,,\qquad 
G_4=\frac{\Mpl^2}{2}\,,\qquad 
G_5=0\,,
\ee
where the bare cosmological constant has been set to zero for simplicity.
Since the action has a symmetry under the simultaneous change~$\gamma_1\to -\gamma_1$ and $\phi\to -\phi$, we assume 
\be
\gamma_1>0\,,
\ee
without loss of generality.
We note that the BH solutions for cubic Galileons 
with the cosmological constant were discussed 
in Ref.~\cite{Babichev:2016fbg} by assuming a time-dependent 
scalar field of the form $\phi=qt+\Phi(r)$, where 
$q$ is a nonvanishing constant. 
Here, we are considering a time-independent scalar field ($q=0$) 
and addressing the linear stability of BHs unexplored in Ref.~\cite{Babichev:2016fbg}.

We exploit the scalar-field equation~(\ref{G3sca}) together with 
the equations of motion~\eqref{back1} and \eqref{back2} for the metric to obtain 
the asymptotic forms of $f$, $h$, and $\phi'$ around 
the BH horizon and at spatial infinity. 
The solutions expanded around $r=r_s$ are given by 
\ba
\frac{f }{f_1}&=& (r-r_s)
+\frac{1}{r_s} \left( \frac{\eta^3 r_s^4}{\gamma_1^2 \Mpl^2} 
-1 \right) (r-r_s)^2+{\cal O} ((r-r_s)^3)\,,
\label{fho}\\
h &=& 
\frac{1}{r_s} (r-r_s)-\frac{1}{r_s^2} 
\left( \frac{3\eta^3 r_s^4}{\gamma_1^2 \Mpl^2} 
+1 \right) (r-r_s)^2+{\cal O} ((r-r_s)^3)\,,
\label{hho}\\
\phi' &=& 
-
\frac{2\eta r_s}{\gamma_1}
-\frac{4 \eta}{\gamma_1}\left( 
\frac{\eta^3 r_s^4}{\gamma_1^2 \Mpl^2} 
-1 \right) (r-r_s)
+{\cal O} ((r-r_s)^2)\,,
\label{phiho}
\ea
where $f_1>0$.
At spatial infinity, we obtain the 
following expansion:
\ba
\frac{f}{f_0} &=&
\frac{\sqrt{6}(-\eta)^{3/2}}
{18 \gamma_1\Mpl }r^2
+1-\frac{r_1}{r}
-\frac{3\sqrt{6}\gamma_1 \Mpl r_1}
{20(-\eta)^{3/2}r^3}
+{\cal O} (r^{-4})\,,\label{fla}\\
h &=& 
\frac{\sqrt{6}(-\eta)^{3/2}}
{18 \gamma_1\Mpl }r^2
+\frac{5}{6}-\frac{r_1}{r}
-\frac{11\sqrt{6}\gamma_1 \Mpl}
{24 (-\eta)^{3/2} r^2} 
+{\cal O} (r^{-3})\,,\label{hla}\\
\phi' &=& 
\frac{\sqrt{6}\Mpl}
{\sqrt{-\eta}r}
-\frac{9 \Mpl^2\gamma_1}
{\eta^2 r^3}
+\frac{9 \Mpl^2\gamma_1 r_1}{\eta^2 r^4}
+\frac{75\sqrt{6} \Mpl^3\gamma_1^2}
{4(-\eta)^{7/2} r^5}
+{\cal O} (r^{-6})\,,\label{phila}
\ea
where $f_0\,(>0)$ and $r_1$ are constants. 
The constant~$f_0$ can be chosen freely 
due to the time reparametrization invariance.
For the existence of this solution, 
we require that 
\be
\eta<0\,.
\label{etane0}
\ee
Note that, as in the case of Schwarzschild-AdS BHs, the metric components $f$ and $h$ are growing functions of $r$ in the region~$r \gg r_s$.
It should also be noted that there exists another branch of solutions where the coefficients of $r^2$ in $f$ and $h$ 
are negative.
We are not interested in this other branch, because a numerical integration from the BH horizon yields the branch of solution with \eqref{fla}--\eqref{phila},
as we shall see below.

In what follows, we verify that the above solution satisfies all the stability conditions for both odd- and even-parity perturbations.
From Eqs.~(\ref{con1}), (\ref{calH}), and (\ref{calF}), 
we have
\be
{\cal G}={\cal H}={\cal F}=\Mpl^2>0\,,
\ee
so that the stability conditions of odd-parity 
perturbations are satisfied.
For even-parity modes, the squared propagation 
speeds of gravitational perturbations 
along the radial and angular directions 
reduce to 
\be
c_{r1,{\rm even}}^2=1\,,\qquad 
c_{\Omega 1}^2=1\,.
\label{csq1_cubic}
\ee
The latter follows from the fact that 
the term~$B_1^2-B_2$ in Eq.~(\ref{cosq}) can 
be factored out in the form~$B_1^2-B_2=B_3^2$, 
where $B_3$ can change its sign depending on 
the coordinate distance~$r$. Then, there are 
the two solutions~$c_{\Omega 1}^2=-B_1+B_3$ 
and $c_{\Omega 2}^2=-B_1-B_3$. 
One of them, which is equivalent to unity, 
corresponds to the propagation speed squared 
in the gravity sector, while the other to 
that of the scalar field. 
Hence, on the static and spherically symmetric 
background the cubic Galileon does not 
modify the propagation speed of gravitational perturbations in comparison to GR 
(analogous to the speed of tensor perturbations on an isotropic cosmological 
background~\cite{KYY,DeFelice:2011bh}).
Around $r=r_s$, the no-ghost condition for the even-parity perturbations 
translates to 
\ba
{\cal K} =
-\frac{2\eta^3 r_s^4}{\gamma_1^2}
+{\cal O}(r-r_s)\,,
\label{Ksm0}
\ea
where we have used the inequality~(\ref{etane0}). 
Since the leading-order contribution to 
Eq.~(\ref{Ksm0}) is positive, 
the ghost is absent.
In the vicinity of the horizon, the 
squared radial and angular propagation speeds 
of the scalar field are given, 
respectively, by 
\ba
c_{r2,{\rm even}}^2 &=& 
1+{\cal O}(r-r_s)\,,\label{cr2sm}\\ 
c_{\Omega 2}^2 &=& 
3+{\cal O}(r-r_s)\,,
\ea
and hence there are no Laplacian 
instabilities around $r=r_s$.
At spatial infinity, the quantities 
${\cal K}$ and $c_{r2,{\rm even}}^2$ 
can be estimated as
\ba
{\cal K} &=&
\frac{\sqrt{6}(-\eta)^{3/2}
\Mpl}{4\gamma_1}r^2+{\cal O}(r)\,,
\label{Kla}\\
c_{r2,{\rm even}}^2 &=&
\frac{(-\eta)^{3/2}}
{\sqrt{6}\gamma_1 \Mpl}r^2+{\cal O}(r)\,.
\label{cr2la}
\ea
Under the condition~(\ref{etane0}), 
the leading-order contributions to 
${\cal K}$ and $c_{r2,{\rm even}}^2$ 
are positive.
For $r \gg r_s$, the quantities~$B_1$ and $B_2$ have the 
following asymptotic behavior:
\be
B_1=-\frac{1}{2}-\frac{r_1}{4r}+{\cal O}(r^{-2})\,,\qquad 
B_2=\frac{r_1}{2r}-\left[ 
\frac{r_1^2}{2}+\frac{5\sqrt{6}
\gamma_1 \Mpl}{4(-\eta)^{3/2}}
\right] 
\frac{1}{r^2}+{\cal O}(r^{-3})\,.
\ee
Then, the second angular propagation speed squared yields
\be
c_{\Omega 2}^2=\frac{r_1}{2r}
+\left[ 
\frac{r_1^2}{2}+\frac{5\sqrt{6}
\gamma_1 \Mpl}{4(-\eta)^{3/2}}
\right] 
\frac{1}{r^2}
+{\cal O}(r^{-3})\,.
\label{csOmega2_cubic}
\ee
Hence, the Laplacian stability along the angular directions 
is ensured for
\be
r_1>0\,.
\label{r1con}
\ee
%

%%%%%%%%%%%%%%%%%%%%%%%%%%%%%%%%
\begin{figure}[ht]
\begin{center}
\includegraphics[height=3.3in,width=3.4in]{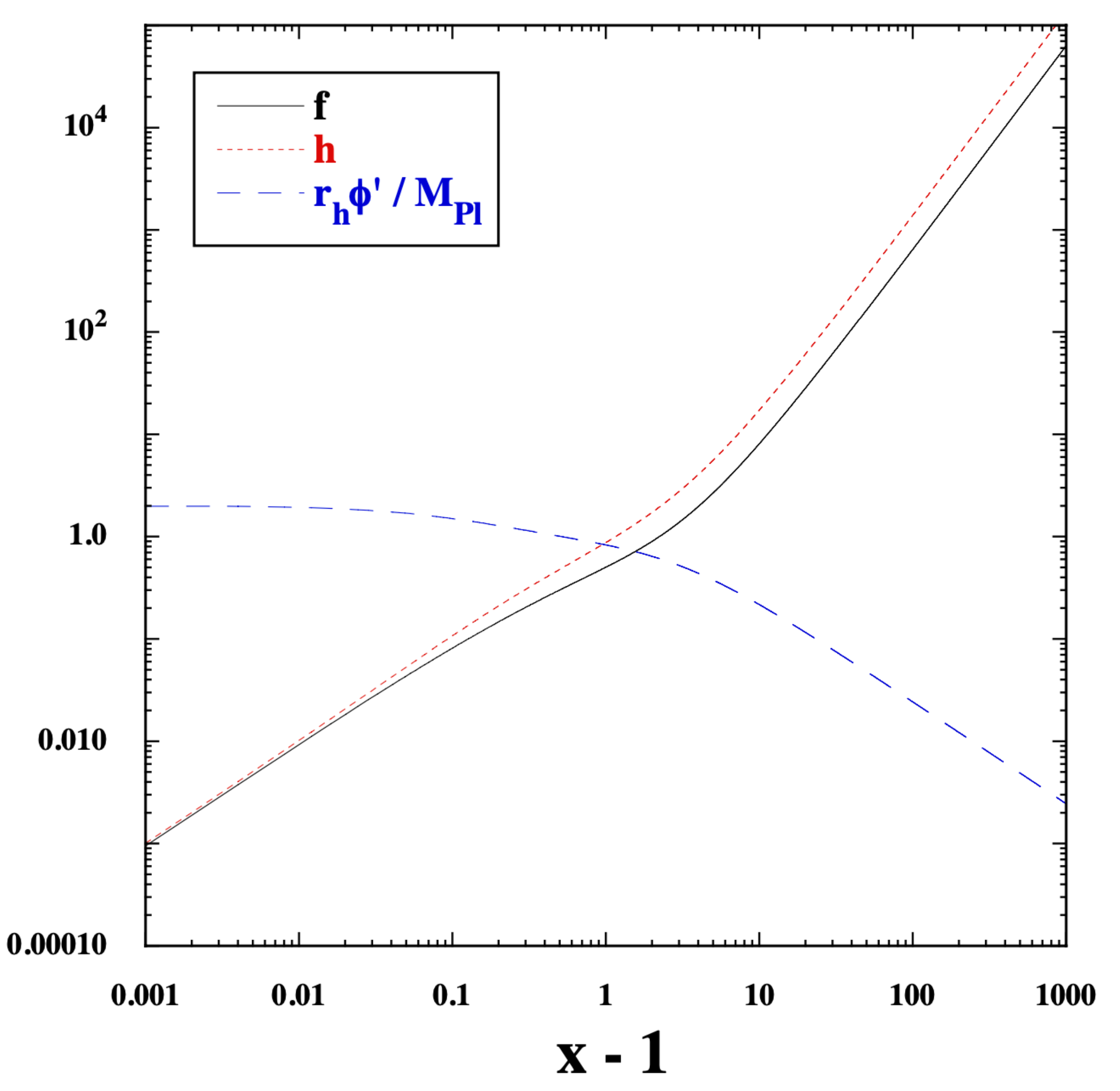}
\includegraphics[height=3.3in,width=3.3in]{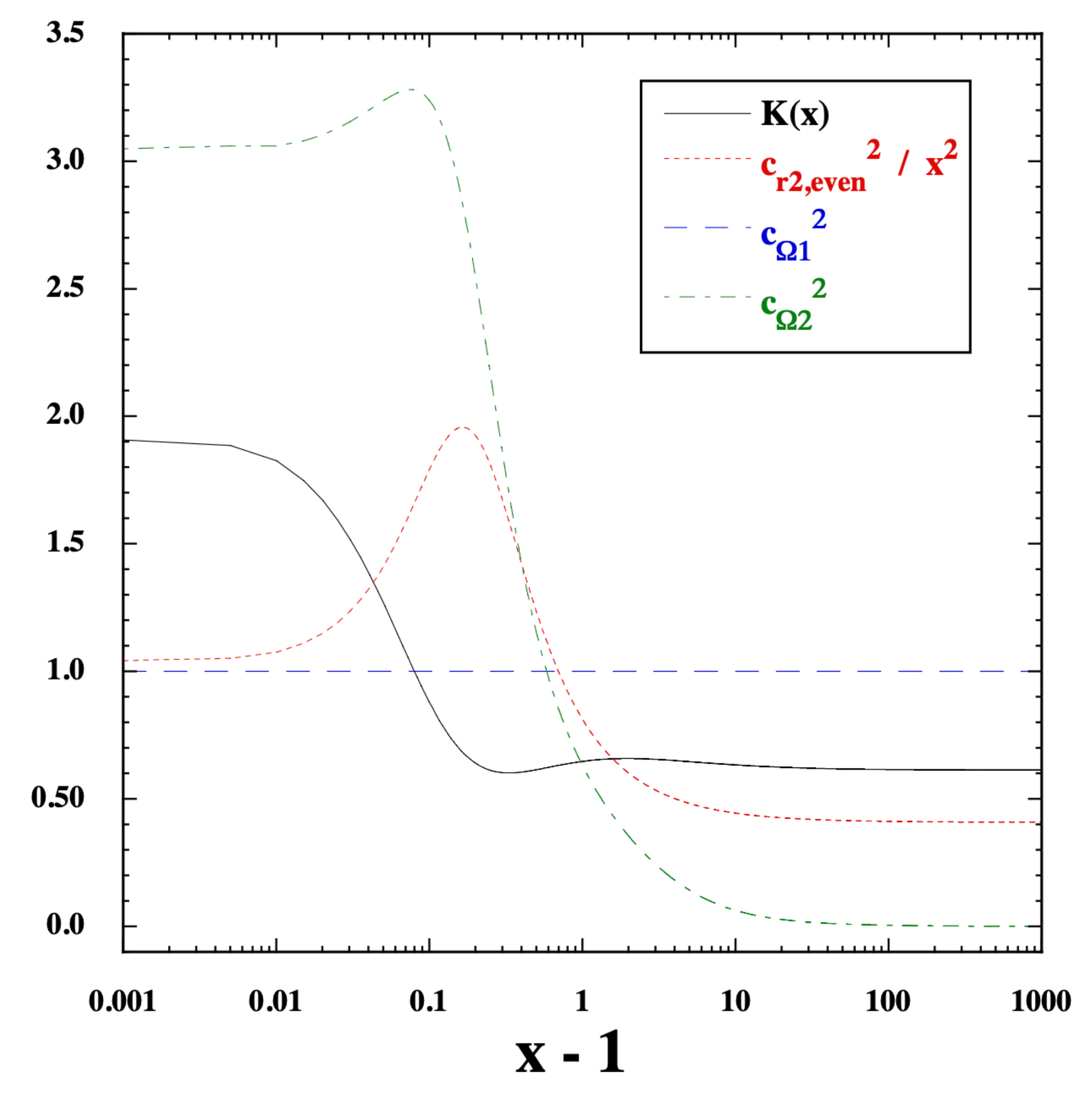}
\end{center}\vspace{-0.5cm}
\renewcommand{\figurename}{FIG}
\renewcommand{\thefigure}{\!\!}
\caption{\label{fig1}
Left: background metric components~$f$, $h$, and the scalar-field derivative~$r_s \phi'/\Mpl$ versus $x-1$ 
for $\hat{\gamma}_1=1$ and $\eta=-1$, 
where $x=r/r_s$.
The background equations of motion are 
integrated outwards from the distance~$x=1+10^{-6}$.
As the boundary conditions around $r=r_s$,  
we adopt the expanded 
solutions~(\ref{fho})--(\ref{phiho}), with $f_1=1/r_s$.
Right: $K(x)={\cal K}/(\Mpl^2x^2)$, 
$c_{r2, {\rm even}}^2/x^2$, 
$c_{\Omega 1}^2$, and $c_{\Omega 2}^2$ 
versus $x-1$ for the same model parameters 
and boundary conditions as those used in the left.}
\end{figure}
%%%%%%%%%%%%%%%%%%%%%%%%%%%%%%%%

The above discussion shows that, 
under the three conditions~$\eta<0$, 
$\gamma_1>0$, and $r_1>0$, 
there are neither ghost nor Laplacian 
instabilities both around the BH horizon 
and at spatial infinity. 
In order to show the existence of 
stable BH solutions, 
we solve the background equations of motion 
outwards from the vicinity of the BH horizon.
For this purpose, we introduce a dimensionless parameter~$\hat{\gamma}_1=\gamma_1 \Mpl/r_s^2$. 
In the left panel of \hyperref[fig1]{Figure},
we plot $f$, $h$, and $r_s \phi'/\Mpl$
versus the coordinate distance from the BH horizon~$x-1$ (where $x=r/r_s$) 
for the coupling~$\hat{\gamma}_1=1$ with $\eta=-1$. 
As estimated from Eqs.~(\ref{fho})--(\ref{phiho}),
the metric components~$f$ and $h$ 
around the horizon linearly 
grow in $r$, with $\phi' \simeq 
{\rm constant}$. 
For $x-1 \gtrsim 10$, $f$ and $h$ 
increase as $f \propto r^2$ and $h \propto r^2$
according to the large-distance solutions~(\ref{fla}) 
and (\ref{hla}), with  
$\phi' \propto 1/r$. 
As $r \to \infty$, the kinetic term~$\eta X$ 
approaches a constant
\be
\eta X \to \frac{(-\eta)^{3/2}\Mpl}
{\sqrt{6}\gamma_1}\,,
\ee
which is positive. This positive asymptotic 
value works as a negative cosmological 
constant, so the leading-order 
metric components at spatial infinity 
are similar to those of Schwarzschild-AdS spacetime. 
As we observe in the left panel of \hyperref[fig1]{Figure}, 
the solutions 
in two asymptotic regimes 
($r \simeq r_s$ and $r \gg r_s$)
are smoothly joined with each other.

In the right panel of 
\hyperref[fig1]{Figure}, 
we plot the quantities associated with 
the stability conditions of even-parity 
perturbations. The variable~$K(x)={\cal K}/(\Mpl^2 x^2)$ is 
positive throughout the horizon exterior, 
with the asymptotic values of ${\cal K}$ 
given by Eqs.~\eqref{Ksm0} and \eqref{Kla}.
The radial propagation speed 
squared~$c_{r2,{\rm even}}^2$ is also 
positive for $r>r_s$, with
asymptotic behaviors~(\ref{cr2sm}) and (\ref{cr2la}).
The angular propagation speed squared~$c_{\Omega1}^2$ 
in the gravity sector is 
unity [see Eq.~\eqref{csq1_cubic}].
Also, \hyperref[fig1]{Figure} clearly shows 
that the positivity of $c_{\Omega2}^2$ holds outside the BH horizon.

While we have shown that the cubic Galileons 
with $\gamma_1>0$ and $\eta<0$ allow the existence of static, 
spherically symmetric, and asymptotically AdS
BH solutions being compatible with
all the linear stability conditions listed in \hyperref[table]{Table},
they are not sufficient to conclude that the BH solutions are physically sensible, 
and further studies would be required.
As $r \to \infty$, $c_{\Omega2}^2$ 
is vanishing, while $c_{r2,{\rm even}}^2$ grows toward infinity,
which might imply the presence of a strong coupling 
problem at the timelike AdS boundary. In order to see 
whether this is the case or not, we need to study 
whether higher-order nonlinear operators dominate 
over those of linear perturbations in the Lagrangian. 
Moreover, the BH stability at the nonlinear level should also be studied,
which is beyond the scope of this paper.

\subsection{Quintic-order positive power-law couplings}
\label{quinsec}

We proceed to study the linear stability of BHs in the presence of quintic-order couplings $G_5(X)$. Let us first consider models with a positive power-law 
coupling given by the functions
\be
G_2=\eta X\,,\qquad G_3=0\,,\qquad 
G_4=\frac{\Mpl^2}{2}\,,\qquad 
G_5=\lambda_p (-X)^p\,,
\label{quinmo}
\ee
where $\lambda_p$ and 
$p\,(>0)$ are constants. 
In this case, the scalar-field equation~(\ref{Jeq}) yields
\be
{\phi'}^{2p-1}=\frac{2^p \eta r^2 f}
{\lambda_p p f' h^p [(2p+1)h-1]}\,.
\label{phiquin}
\ee
Then, the scalar-field derivative diverges 
at the coordinate distance~$r_d$ satisfying 
\be
h(r_d)=\frac{1}{2p+1}\,,
\label{hd}
\ee
provided it exists.
If we impose asymptotic flatness of the spacetime and construct solutions around the BH horizon ($h \simeq 0$) and at large distances ($h \simeq 1$), then the two solutions cannot be smoothly connected without hitting the singular point~\eqref{hd}.
This problem also persists if $h$ 
is a growing function of $r$ toward spatial infinity 
as in the Schwarzschild-AdS spacetime. 
A possible way out is to make the range of $h$ finite as in 
the Schwarzschild-dS spacetime so that it does not reach the singular point~(\ref{hd}).

Let us first consider the model~(\ref{quinmo}) with 
\be
p=1\,,
\ee
for concreteness. 
On using the background equations.~(\ref{back1})--(\ref{back3}), 
the expanded solution at spatial infinity is given by 
\begin{align}
\frac{f}{f_0}&=\pm \left(-\frac{\eta^3}{55566 \Mpl^2\lambda_1^2}\right)^{1/4}r^2+1+{\cal O}(r^{-1})\,, \label{f_inf_quin} \\
h&=\pm \left(-\frac{7 \eta^3}{162 \Mpl^2 
\lambda_1^2} \right)^{1/4}r^2
+\frac{139}{28}+{\cal O}(r^{-1})\,, \label{h_inf_quin}
\end{align}
where $f_0>0$ and the double signs 
are in the same order.
The existence of this solution requires that 
\be
\eta<0\,.
\label{etane}
\ee
For $p=1$, the scalar-field derivative on the BH 
horizon ($r=r_s$) has a nonvanishing value~$\phi'(r_s)=-2\eta r_s^3/\lambda_1$ and hence $X(r_s)=0$, 
where we have used the expansions~(\ref{fexpan}) 
and (\ref{hexpan}).
Around $r=r_s$, the metric 
components are given by 
\ba
\frac{f}{f_1} &=& (r-r_s)
+\frac{\eta^3 r_s^8-\Mpl^2 \lambda_1^2}
{r_s \Mpl^2 \lambda_1^2} (r-r_s)^2
+{\cal O}((r-r_s)^3)\,,\\
h &=& \frac{1}{r_s}(r-r_s)-\frac{3 \eta^3 r_s^8
+\Mpl^2 \lambda_1^2}
{r_s^2 \Mpl^2 \lambda_1^2}(r-r_s)^2
+{\cal O}((r-r_s)^3)\,,
\ea
where $f_1>0$.
If we connect the above expanded solutions and require that $h$ does not reach the singular point~\eqref{hd}, then we should at least choose the minus sign in Eqs.~\eqref{f_inf_quin} and \eqref{h_inf_quin}.
On using the expanded solution around $r=r_s$, the quantity 
associated with the no-ghost condition of 
even-parity perturbations reads
\begin{align}
{\cal K}=-\frac{2\eta^3 r_s^8}{\lambda_1^2}
+{\cal O}(r-r_s)\,,
\label{Ksm}
\end{align}
whose leading-order term is positive 
under the inequality~(\ref{etane}). 
However, we have 
\be
B_2=-3+\frac{2\eta^3 r_s^8}{\Mpl^2 \lambda_1^2}\,,
\ee
which is negative.
This means that, even if the two solutions around 
$r=r_s$ and $r \to \infty$ are connected without reaching the 
singular point~(\ref{hd}),
the Laplacian instability of even-parity 
perturbations is present around 
the BH horizon.

Next, let us consider the case 
with $p>1$.
Assuming the existence of a BH horizon, from Eq.~\eqref{phiquin}, the leading term of the scalar-field derivative around $r=r_s$ is given by $\phi'\propto \eta^{1/(2p-1)}(r-r_s)^{-(p-1)/(2p-1)}$.
Then, one can verify that the background equations~\eqref{back1} and \eqref{back2} are consistently satisfied around $r=r_s$ only when $\eta=0$, for which we have $\phi'=0$ everywhere.
Hence, for $p>1$, a nontrivial scalar-field profile is not present even at the background level.

\subsection{Gauss-Bonnet couplings}
\label{Gausssec}

Finally, we consider the case with the scalar field linearly 
coupled to the Gauss-Bonnet curvature invariant~$R_{\rm GB}^2 \equiv R^2-4R^{\alpha\beta}R_{\alpha\beta} +R^{\alpha\beta\mu\nu}R_{\alpha\beta\mu\nu}$, which is described by the action
\be
\label{scalar_GB2}
{\cal S}=\int {\rm d}^4x
\sqrt{-g} \left[
\frac{\Mpl^2}{2}R
+\eta X
+\alpha_{\rm GB}\phi
R_{\rm GB}^2
\right]\,,
\ee
where $\alpha_{\rm GB}$ is a coupling constant.
This theory can be accommodated in the framework of 
shift-symmetric Horndeski theories 
with the following choice of the coupling functions~\cite{KYY}:
\be
\label{scalar_GB}
G_2=\eta X,
\qquad 
G_3=0,
\qquad
G_4=\frac{\Mpl^2}{2},
\qquad 
G_5=-4\alpha_{\rm GB} \ln |X|\,.
\ee
In this theory, the background 
equations~(\ref{back1}) and (\ref{back2}) yield
\ba
&&
2\Mpl^2
-2 h' \left( \Mpl^2r
+4\alpha_{\rm GB}\phi' \right)
+16\alpha_{\rm GB} h^2 \phi''
-h \left(2\Mpl^2
-24\alpha_{\rm GB} h'\phi'
+\eta r^2\phi'^2+16 \alpha_{\rm GB}\phi''
\right)
=0\,,\label{eom_gb1}\\
&&
2h f'
\left[
\Mpl^2r
+4 \alpha_{\rm GB}(1-3h) \phi'
\right]
-f \left[
2\Mpl^2
-h (2\Mpl^2-\eta r^2\phi'^2)
\right]=0\,.
\label{eom_gb2}
\ea
The radial component of the current~$J^{\mu}$ reduces to
\be
J^r=h
\left[
\eta \phi'
+4\alpha_{\rm GB} \frac{(h-1)f'}{r^2 f}
\right]\,,
\label{JrGB}
\ee
which obeys Eq.~(\ref{Jreq}). 
The general solution to 
Eq.~(\ref{Jreq}) is 
given by Eq.~(\ref{Jrso}), so that 
\be
h \left[
\eta \phi'
+4\alpha_{\rm GB} \frac{(h-1)f'}{r^2 f}
\right]=\frac{Q}{r^2} \sqrt{\frac{h}{f}}\,.
\label{JeqGB}
\ee
As we discussed in Sec.~\ref{scasec}, 
the finiteness of the current squared~$J^2$ 
requires that $Q=0$, in which case $J^r=0$. 
For $Q \neq 0$, $J^2$ diverges on the horizon. 
In this latter case, it was argued that 
hairy BH solutions with a nonvanishing scalar-field 
derivative are present. In spite of the divergence 
of $J^2$, the components of the energy-momentum 
tensor and curvature invariants remain finite. 
In Ref.~\cite{Creminelli:2020lxn}, it was argued that 
asymptotically flat BHs in shift-symmetric scalar-tensor theories without ghost degrees 
of freedom can have nontrivial scalar hair only 
in the presence of the Gauss-Bonnet coupling~$\alpha_{\rm GB}\phi R_{\rm GB}^2$.

In the following, we will consider the two 
cases~$Q=0$ and $Q \neq 0$ in turn.

\subsubsection{\texorpdfstring{$Q=0$}{Q=0}}

When $Q=0$, Eq.~(\ref{JeqGB}) gives 
\be
\phi'=-\frac{4\alpha_{\rm GB}(h-1)f'}
{\eta r^2 f}\,.
\ee
On using the expansions~(\ref{fexpan}) 
and (\ref{hexpan}) around the BH horizon~$r=r_s$, 
the scalar-field derivative 
has the following dependence:
\be
\phi'=\frac{4\alpha_{\rm GB}}
{\eta r_s^2 (r-r_s)}
+{\cal O}((r-r_s)^0)\,.
\ee
This means that, in the vicinity of the BH horizon, 
the left-hand side of Eq.~(\ref{eom_gb1})
behaves as
\ba
\frac{16
h_1\alpha_{\rm GB}^2}{\eta r_s^2(r-r_s)}
+{\cal O}((r-r_s)^0)\,,
\ea
which does not vanish. 
Hence, for $Q=0$, we do not have a 
BH solution endowed with scalar hair. 
This conclusion agrees with the 
one reached in Ref.~\cite{Babichev:2017guv}. 

\subsubsection{\texorpdfstring{$Q\ne 0$}{Q!=0}}

For $Q \neq 0$, it is possible to realize a solution 
with a finite value of $\phi'$ on the BH horizon. 
Applying the expansions~(\ref{fexpan}) 
and (\ref{hexpan}) of $f$ and $h$ to Eq.~(\ref{JeqGB}), 
the divergence of $\phi'$ 
can be avoided for 
\be
Q=-4\alpha_{\rm GB} \sqrt{f_1 h_1}\,.
\label{QGB}
\ee
Then, the scalar-field derivative at $r=r_s$ takes 
a finite constant value, 
\be
\phi'(r_s)=
-\frac{2\alpha_{\rm GB}
[(2h_1^2-h_2)f_1-3h_1 f_2]}
{\eta f_1 h_1 r_s^2}\,,
\label{phirhGB}
\ee
so that $X(r_s)=0$. 
The hairy BH solution in 
Refs.~\cite{Sotiriou:2013qea,Sotiriou:2014pfa}, 
whose existence was studied in both perturbative 
and numerical approaches, corresponds to 
the nonvanishing value of $Q$ given 
by Eq.~(\ref{QGB}).
Here, we follow the perturbative approach 
valid in the regime of small Gauss-Bonnet couplings. 
Substituting Eq.~(\ref{QGB}) into Eq.~(\ref{JeqGB}) 
and taking the limit~$\alpha_{\rm GB} \to 0$, we 
obtain the no-hair solution~$\phi'=0$.
In this limit, Eqs.~(\ref{eom_gb1}) and 
(\ref{eom_gb2}) show that the corresponding 
background geometry is the 
Schwarzschild spacetime.
For small Gauss-Bonnet couplings, 
we expand $f$, $h$, and $\phi'$ in terms of the 
dimensionless constant~${\hat \alpha}_{\rm GB} \equiv 
\alpha_{\rm GB}/(m^2 \Mpl)$, as 
\ba
\label{sch_pert}
f(r)
&= &
\left(1-\frac{2m}{r}\right)
\left[
1+
\sum_{j=1}^\infty
\hat{f}_j(r)
({\hat \alpha}_{\rm GB})^j
\right]^2,
\qquad
h(r)
=
\left(1-\frac{2m}{r}\right)
\left[
1+
\sum_{j=1}^\infty
\hat{h}_j(r)
\left(\hat{\alpha}_{\rm GB}\right)^j
\right]^{-2},
\nonumber
\\
\phi'(r)
&=&
\sum_{j=1}^\infty
\phi'_j(r)
\left({\hat \alpha}_{\rm GB}\right)^j\,,
\ea
where $m$ is constant 
and $\hat{f}_j$, $\hat{h}_j$, $\phi_j'$ are 
functions of $r$.
We substitute the ansatz~\eqref{sch_pert}
into Eqs.~(\ref{eom_gb1}), (\ref{eom_gb2}), 
(\ref{Jreq}) with (\ref{JrGB})
and solve them at each order of 
$\hat{\alpha}_{\rm GB}$
up to the second order ($j=2$).
Although the position of the BH horizon corresponds 
to $r=2m$, as we will see below, 
the Arnowitt-Deser-Misner~(ADM) mass is 
different from $2m$ due to the contribution 
of Gauss-Bonnet couplings.

At first order in $\hat{\alpha}_{\rm GB}$, 
the resulting solutions are expressed 
in the forms
\be
\hat{f}_1(r)=
-\frac{C_2}{r-2m}+C_3\,,
\qquad
\hat{h}_1(r)
=\frac{C_2}{r-2m}\,,
\qquad
\phi'_1(r)
=
\frac{16m^4 \Mpl+r^3\eta C_4}{r^4(r-2m)\eta}\,,
\ee
where $C_2$, $C_3$, and $C_4$ are integration constants.
We set $C_3=0$ by a suitable time reparametrization.
We also impose the regularity of 
perturbative solutions at $r=2m$,
which yields $C_2=0$ and $C_4=-2m\Mpl/\eta$.
Thus, the first-order solutions 
are given by 
\be
\label{pert1}
\hat{f}_1(r)=\hat{h}_1(r)=0\,,
\qquad
\phi'_1(r)=-\frac{2m \Mpl(r^2+2mr+4m^2)}
{\eta r^4}\,.
\ee
Up to this order, the field derivative on the 
horizon is $\phi'(r_s)=-3\alpha_{\rm GB}
/(2\eta m^3)$. Indeed, substituting the 
zeroth-order metric components~$f_1=h_1=1/(2m)$ 
and $f_2=h_2=-1/(4m^2)$ with $r_s=2m$ into 
Eq.~(\ref{phirhGB}), we obtain the 
same value of $\phi'(r_s)$ 
at first order in $\alpha_{\rm GB}$.

At the second order in $\alpha_{\rm GB}$, 
the solution is given by
\be
\label{pert2}
\begin{split}
\hat{f}_2(r)
&=
\frac{m\left(1600m^5+416m^4r-56m^3r^2
-548m^2r^3-294m r^4-147r^5\right)}{120\eta r^6}\,,
\\
\hat{h}_2(r)
&=
-\frac{m\left(7360m^5+3488m^4r+1624m^3r^2-228m^2r^3-174m r^4-147r^5\right)}{120 \eta r^6}\,,
\\
\phi'_2(r)
&=
0\,,
\end{split}
\ee
where the integration constants have been fixed 
by using the time reparametrization invariance 
for $\hat{f}_2(r)$ and by imposing the regularities 
of $\hat{h}_2(r)$ and $\phi'_2(r)$ at $r=2m$.
At large distances ($r \gg 2m$), 
$f$, $h$, and $\phi'$ up to
${\cal O}(\alpha_{\rm GB}^2)$ 
are expressed in the forms 
\ba
f(r)
&=&
1
-\frac{2M}{r}
+
{\cal O}
\left(
r^{-2}
\right)\,,
\qquad
h(r)
=
1
-\frac{2M}{r}
+
{\cal O}
\left(
r^{-2}
\right)\,,
\qquad
\phi'(r)
=
-\frac{C}{r^2}
+
{\cal O}
\left(
r^{-3}
\right)\,,
\ea
where 
\be
M=
m+
\frac{49\alpha_{\rm GB}^2}{40m^3\Mpl^2\eta}\,,
\qquad 
C=
\frac{2\alpha_{\rm GB}}{m\eta}\,.
\ee
Here, $M$ and $C$ correspond to the ADM mass 
and scalar charge, 
respectively~\cite{Sotiriou:2013qea,Sotiriou:2014pfa}. 
Since both $M$ and $C$ are determined solely by $m$,
the scalar charge~$C$ is of secondary type. 
Substituting the leading-order metric 
components~$f_1=h_1=1/(2m)$ into Eq.~(\ref{QGB}), we obtain 
$Q=-2\alpha_{\rm GB}/m$ and hence $Q=-\eta C$.

In Ref.~\cite{Sotiriou:2014pfa}, it was shown 
that the perturbative solutions~\eqref{sch_pert} with Eqs.~\eqref{pert1} and 
\eqref{pert2} exhibit 
very good agreement with 
full numerical results.
Moreover, since the numerical BH solution could 
be constructed only for smaller couplings 
$|\hat{\alpha}_{\rm GB}| \ll  {\cal O}(0.1)$~\cite{Sotiriou:2013qea,Sotiriou:2014pfa},
the perturbative solutions~\eqref{sch_pert} 
with Eqs.~\eqref{pert1} and \eqref{pert2} should be 
valid for all the 
coupling regimes in which the BH solutions exist.
Also, by repeating the above procedure, one can compute the perturbative expansion of the solution to an arbitrary order.
One can verify that only the even-order terms ($j=2,4,6,\cdots$) 
of metric components and the odd-order ($j=1,3,5,\cdots$) 
terms of scalar field are nontrivial.
On using these solutions, 
the radial component of $J^{\mu}$, up to 
the order of $\hat{\alpha}_{\rm GB}$, 
is given by
\be
r^2
\sqrt{\frac{f}{h}}J^r
=
-2m \Mpl\hat{\alpha}_{\rm GB}
+{\cal O} \left(
\hat{\alpha}
_{\rm GB}^3 \right)\,.
\label{rsJr}
\ee
Indeed, the leading-order term on the right-hand side of Eq.~(\ref{rsJr}) also follows by 
substituting $Q=-2\alpha_{\rm GB}/m$ 
into Eq.~(\ref{Jrso}).
The nonvanishing value of $J^r$ gives rise to 
a divergent norm of the Noether 
current~$J^2=(J^r)^2/h$ on the horizon ($h=0$). 
As we already mentioned, the components of 
the energy-momentum tensor and curvature invariants remain finite. 
As stated in Ref.~\cite{Creminelli:2020lxn}, the divergence 
of $J^2$ does not give rise to pathologies on the 
BH properties at least in this case.
According to these arguments, the hairy BHs  
discussed above should be dealt as 
physical solutions.

We then study stability of the hairy solution~\eqref{sch_pert} 
with Eqs.~\eqref{pert1} and \eqref{pert2}
against odd- and even-parity perturbations.
The quantities relevant to stability 
in the odd-parity sector yield
\ba
{\cal F}
&=&
\Mpl^2
+\Mpl^2
\frac{16(36-2\hr-\hr^2-2\hr^3)}
{\eta \hr^6}
\hat{\alpha}_{\rm GB}^2
+{\cal O}
\left(\hat{\alpha}_{\rm GB}^4 \right)\,,\\
{\cal G}
&=&
\Mpl^2
+
\Mpl^2
\frac{16(4+2\hr +\hr^2)}{\eta \hr^6}
\hat{\alpha}_{\rm GB}^2
+{\cal O}
\left(\hat{\alpha}_{\rm GB}^4 \right)\,,\\
{\cal H}
&=&
\Mpl^2
+
\Mpl^2
\frac{16(\hr^3-8)}{\eta \hr^6}
\hat{\alpha}_{\rm GB}^2
+{\cal O}
\left(\hat{\alpha}_{\rm GB}^4 \right)\,,
\ea
where we have introduced the dimensionless radial 
coordinate~$\hr \equiv r/m$. 
Provided that $|\hat{\alpha}_{\rm GB}|/\sqrt{|\eta|}\ll 1$,
the terms of order~$\hat{\alpha}_{\rm GB}^2$
in ${\cal F}$, ${\cal G}$, and ${\cal H}$ are 
suppressed relative to the leading-order contribution~$\Mpl^2$ throughout the horizon 
exterior ($2<\hr<\infty$).
This means that, under the validity of the expansion 
with respect to $\hat{\alpha}_{\rm GB}$, i.e.,
\be
\frac{|\alpha_{\rm GB}|}
{m^2 \Mpl \sqrt{|\eta|}}
\ll 1\,,
\label{small_c}
\ee
the stability conditions against odd-parity perturbations
(${\cal F}>0$, ${\cal G}>0$, and ${\cal H}>0$)
are satisfied.

The quantity associated with the no-ghost condition 
of even-parity perturbations yields
\be
{\cal K}=
\Mpl^2
\frac{2(4+2\hr+\hr^2)^2}{\eta \hr^6}
{\hat\alpha}_{\rm GB}^2
+{\cal O}
\left( \hat{\alpha}_{\rm GB}^4 \right)\,.
\ee
Thus, ${\cal K}>0$ is satisfied as long as 
\be
\eta>0\,.
\ee
The squared sound speeds of even-parity 
perturbations along the radial direction
are given by
\ba
c_{r1,{\rm even}}^2
&=&
1+\frac{32(\hr-2)(\hr^2+3\hr+8)}{\eta \hr^6}
\hat{\alpha}_{\rm GB}^2
+{\cal O}
\left( \hat{\alpha}_{\rm GB}^4 \right)\,,\\
\label{cr22}
c_{r2,{\rm even}}^2
&=&
1+
{\cal O}
\left( \hat{\alpha}_{\rm GB}^4 \right)\,.
\ea
We note that, in order to obtain Eq.~\eqref{cr22}, 
one has to solve the background metric components and 
scalar field~\eqref{sch_pert} up to the quartic and cubic 
orders of $\hat{\alpha}_{\rm GB}$ (i.e., $j=4$ and $j=3$)
respectively, though we do not show the explicit forms 
of these higher-order corrections.
The squared sound speeds of 
even-parity perturbations along 
the angular directions are
\be
c_{\Omega,\pm}^2
=
1 \pm \frac{24} 
{\hr^3} 
\sqrt{\frac{2}{\eta}}
|{\hat\alpha}_{\rm GB}|
+\frac{8(84+2\hr+\hr^2-\hr^3)}
{\eta \hr^6}
{\hat\alpha}_{\rm GB}^2
+{\cal O}
\left( \hat{\alpha}_{\rm GB}^3 \right)\,.
\ee
Under the condition~(\ref{small_c}), the corrections to 
the sound speeds arising from the Gauss-Bonnet coupling 
are much smaller than unity throughout the horizon exterior, 
so that 
$c_{r1,{\rm even}}^2\simeq 1$,
$c_{r2,{\rm even}}^2\simeq 1$,
and 
$c_{\Omega,\pm}^2 \simeq 1$.
Thus, in the small-coupling regime, the hairy BH solutions
discussed above suffer from neither ghost 
nor Laplacian instabilities.

%%%%%%%%%%%%%%%%%%%%%%%%%%%%%%%%%%%%%%%%%%
\section{Conclusions}
\label{concludesec}
%%%%%%%%%%%%%%%%%%%%%%%%%%%%%%%%%%%%%%%%%%

In this paper, we addressed the stability of 
hairy BHs in shift-symmetric Horndeski theories 
against linear perturbations on a static and 
spherically symmetric background. 
We assumed that the background 
scalar field is time independent, 
but did not necessarily
impose the asymptotic flatness at spatial infinity. 
Moreover, we allowed the possibility that the 
coupling functions are nonanalytic functions of $X$.
In such cases, the no-hair theorem of BHs in 
shift-symmetric Horndeski theories established in 
Ref.~\cite{Hui:2012qt} 
can be avoided, so that there are several classes 
of hairy BH solutions.
If we require that the norm of the Noether 
current~$J^{\mu}$ associated with the shift symmetry of the scalar field is finite on the horizon, 
the radial current component~$J^r=h \phi' {\cal J}$
vanishes everywhere. 
Provided ${\cal J}$ is finite in the limit 
of $\phi'\to 0$, there is 
a branch of nonvanishing field derivative 
($\phi' \neq 0$) besides a trivial GR solution ($\phi'=0$).

The linear perturbations 
about the static and spherically 
symmetric background can be decomposed into those 
of the odd- and even-parity sectors. 
We clarified the conditions for the absence of ghosts and Laplacian instabilities along the radial and angular directions, which are summarized in \hyperref[table]{Table} at the end of Sec.~\ref{stasec}.
In Sec.~\ref{stealthsec}, we 
applied these conditions to the GR branch~$\phi'=0$ and showed that 
there are no ghost/Laplacian instabilities under
the conditions~\eqref{con1_stealth} and \eqref{con2_stealth}.

In Sec.~\ref{refsec}, we studied the linear stability of hairy BH 
solutions in reflection-symmetric theories containing two arbitrary functions~$G_2(X)$ and $G_4(X)$. 
We employed the expansions~\eqref{fexpan}--\eqref{Xexpan} around the BH horizon to see whether all the conditions in \hyperref[table]{Table} can be consistently satisfied.
As we see in Eq.~\eqref{FKB2_general}, the product of three quantities~${\cal F}$, ${\cal K}$, and $B_2$ is negative around the horizon so long as the conditions in Eq.~\eqref{Xhcon} are satisfied.
Therefore, three of the stability conditions, i.e., ${\cal F}>0$, ${\cal K}>0$, and $B_2>0$ cannot be satisfied at the same time in general.
For instance, even if we require the absence of Laplacian instabilities along the radial direction in the odd modes (${\cal F}>0$) 
and of ghosts in the even modes (${\cal K}>0$), we have $B_2<0$, and correspondingly there would be Laplacian instabilities along the angular directions in the even modes.
In this sense, the instability found here is generic and almost all hairy BHs in the shift- and reflection-symmetric Horndeski theories are shown to be linearly unstable.
%%%%%
We note that a BH solution with $X<0$ at the BH horizon cannot be extended to the interior of the BH horizon 
and can be defined only in the domain outside the horizon where the character of the scalar field is spacelike,
as otherwise the coordinate invariant~$X$ would have an unphysical jump across the horizon.
Thus, BH solutions with $X\neq 0$ at the horizon suffer from 
instabilities only in the domain outside the BH horizon
where the character of the scalar field is spacelike.
Similarly,
a static and spherically symmetric solution with $X<0$ 
at the cosmological horizon (if it exists)
could not be extended to the exterior of the cosmological horizon.
%%%%%
As specific examples, this generic instability is present 
for exact BH solutions 
in theories with $G_4\supset X$ and in those with $G_4\supset (-X)^{1/2}$.
We also showed that, in k-essence
theories given by the Lagrangian~${\cal L}=G_2(X)+(\Mpl^2/2)R$, 
the corresponding BH solution is plagued 
by a strong coupling problem.

In Sec.~\ref{cuqusec}, we investigated the linear stability 
of BH solutions in nonreflection-symmetric 
theories containing either 
$G_3(X)$ or $G_5(X)$. 
For cubic Galileons characterized by the function~$G_3\propto X$, we found the existence of 
nonasymptotically flat solutions satisfying all the conditions for the absence of ghosts/Laplacian instabilities.
In this case, the angular 
propagation speed squared 
of scalar-field perturbations 
is vanishing at spatial infinity. In order to see whether this induces a 
strong coupling problem or not, the analysis of 
nonlinear perturbations is required.
For quintic power-law 
couplings~$G_5\propto (-X)^p$
with $p \geq 1$, it  
turned out to be difficult to realize stable BH solutions with nontrivial scalar hair. 
In the case of the scalar field linearly coupled to the Gauss-Bonnet curvature invariant, which is described by the 
coupling~$G_5\propto \alpha_{\rm GB} \ln |X|$,
we showed that asymptotically flat BH solutions constructed perturbatively with respect to a small coupling~$\alpha_{\rm GB}$ are free of ghosts/Laplacian instabilities.
This is a specific example in which 
asymptotically flat BH solutions 
with nontrivial scalar hair can be 
realized in the framework of 
shift-symmetric Horndeski theories.

We thus narrowed down the range of viable BH solutions 
by scrutinizing 
the linear stability conditions 
including the angular propagation of even-parity perturbations. 
It will be of interest to study 
further whether what kinds of hairy BHs 
survive as stable solutions 
in full Horndeski theories
or in a broader framework of degenerate higher-order scalar-tensor theories~\cite{Langlois:2015cwa,Crisostomi:2016czh,BenAchour:2016fzp,Takahashi:2017pje,Langlois:2018jdg,Takahashi:2021ttd}.
This issue is left for future work.

While we have focused on the linear stability of 
static and spherically symmetric solutions,
we expect that our results should still be valid for 
more general BH solutions in Horndeski theories
where the deviation from staticity and/or 
spherical symmetry is small, for instance
for slowly rotating solutions (see e.g., \cite{Pani:2009wy,Pani:2011gy,Sotiriou:2013qea,Maselli:2015tta,Maselli:2015yva,Cisterna:2015uya}) or 
for static BHs with the small deviation from 
spherical symmetry (if they exist).
In such cases, 
we can treat the deviation from staticity and/or spherical symmetry
as small perturbations to our case,
which would not modify our main results significantly.
%%%%%%%%
On the other hand, the perturbative treatment no longer 
applies to BH solutions with the large deviation from 
staticity and/or spherical symmetry,
e.g., rapidly rotating BHs 
which have been explored in the context of Einstein-scalar-Gauss-Bonnet theory (see e.g., \cite{Kleihaus:2011tg,Kleihaus:2015aje,Collodel:2019kkx,Delgado:2020rev,Herdeiro:2020wei,Berti:2020kgk})
but (to our knowledge) not fully yet in other 
Horndeski classes.
In this case, we have to develop a new theoretical scheme
of BH perturbations to 
clarify whether results similar to the case of 
static and spherically symmetric BHs hold.
%%%%%%
Nevertheless, we speculate that,
irrespective of the presence of rotation and/or 
deviation from the spherical symmetry, 
BH solutions with the nonvanishing constant kinetic term on the (either BH or cosmological) horizon 
cannot be extended across the horizon, as otherwise they admit an unphysical jump of $X$ 
and suffer from similar ghost/Laplacian instabilities 
discussed in this paper.
We hope to come back to these issues in future.

%%%%%%%%%%%%%%%%%%
\section*{Acknowledgments}
%%%%%%%%%%%%%%%%%%

MM was supported by the Portuguese national fund 
through the Funda\c{c}\~{a}o para a Ci\^encia e a Tecnologia (FCT) in the scope of the framework of the Decree-Law 57/2016 
of August 29, changed by Law 57/2017 of July 19,
and the Centro de Astrof\'{\i}sica e Gravita\c c\~ao (CENTRA) through the Project~No.~UIDB/00099/2020.
KT was supported by JSPS (Japan Society for the Promotion of Science) 
KAKENHI Grant No.~JP21J00695.
ST was supported by the Grant-in-Aid for Scientific Research 
Fund of the JSPS No.~19K03854.

%%%%

\appendix*

\renewcommand{\theequation}{A\arabic{equation}}
\setcounter{equation}{0}

\section{Explicit form of the coefficients}
\label{App}

%\mm{[New section]}

Here, we define the functions appearing in the main text.
For the background, the quantities~$A_1,\cdots,A_5$ in 
Eqs.~\eqref{back1}--\eqref{back3}
are given by  
\be
\begin{split}
&
A_1=-h^2 G_{3,X}\phi'^2\,, \\
&
A_2=4 h^2 \phi' \left( h G_{4,XX} \phi'^2-G_{4,X} \right)\,, \\
&
A_3=h^2G_{5,X} ( 3 h-1) \phi'^2-h^4G_{5,XX} \phi'^4\,, \\
&
A_4=2h^2 G_{4,XX} \phi'^4-4h G_{4,X}  \phi'^2-2 G_4\,, \\
&
A_5=-\frac12 \left[G_{5,XX} h^3{\phi'}^{5}
- h( 5 h-1 )G_{5,X} \phi'^3\right]\,.
\end{split}
\label{A1_A5}
\ee
For the perturbations, the quantity~$a_1$ 
in Eq.~\eqref{p1mu} is defined by 
\ba
\label{a1}
a_1&=&\frac{\sqrt{fh}}{2} \left[ 
G_{3,X} h\phi'^2 r^2+4 (G_{4,X}-G_{4,XX}h \phi'^2)h \phi' r
+G_{5,XX} h^3\phi'^4-G_{5,X} h ( 3 h-1 ) \phi'^2 \right]\,. 
\ea
The definitions of $c_2$ and $c_4$ 
in Eq.~\eqref{cr2} are
\begin{align}
\label{c2}
c_2=&\phi'\sqrt{fh} \biggl\{ \frac{1}{2} (G_{2,X}-G_{2,XX}h \phi'^2) r^2-\frac{(rf'+4f)h r\phi'}{4f} (3G_{3,X}-G_{3,XX}h \phi'^2) 
\nonumber \\
&\qquad\qquad -\frac{h(rf'+f)}{f} ( 3G_{4,X}-6G_{4,XX}h\phi'^2+G_{4,XXX}h^2\phi'^4 )
+G_{4,X}-G_{4,XX} h\phi'^2
\nonumber \\
&\qquad\qquad +\frac{f'h\phi'}{4f}\left[
3 G_{5,X}  ( 5 h-1 ) -G_{5,XX} h ( 10 h-1 ) \phi'^2 + G_{5,XXX} h^3\phi'^4 \right] \biggr\}\,, 
\\
\label{c4}
c_4=&\frac{\phi'}{4} \sqrt{\frac{h}{f}} 
\left[2G_{3,X}f\phi'+\frac{2(rf'+2f)}{r}(G_{4,X}-G_{4,XX}h\phi'^2)-\frac{f'h\phi'}{r}(3G_{5,X}-G_{5,XX}h\phi'^2)\right]\,. 
\end{align}
The functions~$B_1$ and $B_2$ in Eq.~\eqref{cosq}
are given by
\ba
\label{B1B2}
&&
B_1=
\frac {r^3\sqrt {f h} {\cal H} [ 4 h ( \phi' a_1+r\sqrt{fh} {\cal H}) 
\beta_1+\beta_2-4 \phi' a_1 \beta_3] 
-2 fh {\cal G}  [ r \sqrt{fh}( 2 {\cal P}_1-{\cal F}){\cal H}  
( 2\phi' a_1+r\sqrt{fh} {\cal H} )+2\phi'^2a_1^{2}{\cal P}_1 ] }
{4f h ( 2 {\cal P}_1-{\cal F} ){\cal H}
(\phi' a_1+r\sqrt{fh} {\cal H})^2}\,,
\label{B1def}
\notag\\\\
&&
B_2=
-r^2{\frac {r^2h \beta_1 [ 2 fh {\cal F} {\cal G} ( \phi' a_1+r \sqrt{fh}{\cal H} ) 
+r^2\beta_2 ] -{r}^{4}\beta_2 \beta_3
-fh{\cal F} {\cal G}  ( \phi' fh {\cal F} {\cal G}a_1 +2 r^3 \sqrt{fh} {\cal H} \beta_3 ) }
{ fh\phi' a_1 ( 2 {\cal P}_1-{\cal F} ){\cal F}  ( \phi' a_1+r \sqrt{fh}{\cal H} ) ^{2}}}\,,
\label{B2def}
\ea
with 
\ba
\beta_1&=&\frac12 \phi'^2 \sqrt{fh} {\cal H}e_4 
-\phi' \left(\sqrt{fh}{\cal H} \right)' c_4 
+ \frac{\sqrt{fh}}{2}\left[ \left( {\frac {f'}{f}}+{\frac {h'}{h}}-\frac{2}{r} \right) {\cal H}
+{\frac {2{\cal F}}{r}} \right] \phi' c_4+{\frac {f{\cal F} {\cal G}}{2r^2}}\,,\\
\beta_2&=& \left[ \frac{\sqrt{fh}{\cal F}}{r^2} \left( 2 hr\phi'^2c_4
+\frac{r \phi' f' \sqrt{h}}{2\sqrt{f}}{\cal H}-\phi' \sqrt{fh}{\cal G} \right)
-\frac{\phi' fh {\cal G}{\cal H}}{r} \left( \frac{{\cal G}'}{{\cal G}}
-\frac{{\cal H}'}{{\cal H}}+\frac{f'}{2f}-\frac{1}{r} \right) \right]a_1 
-\frac{2}{r} (fh)^{3/2}{\cal F}{\cal G}{\cal H}\,, \qquad\,\,\\
\beta_3&=& \frac{\sqrt{fh}{\cal H}}{2}\phi'  
\left( hc_4'+\frac12 h' c_4-\frac{d_3}{2} \right) 
-\frac{\sqrt{fh}}{2} \left( \frac{\cal H}{r}+{\cal H}' \right) 
\left( 2 h \phi'c_4+\frac{\sqrt{fh}{\cal G}}{2r}
+\frac{f'\sqrt{h}{\cal H}}{4\sqrt{f}} \right)\notag\\
&&
+{\frac {\sqrt {fh}{\cal F}}{4r} \left(  2 h \phi'c_4
+\frac{3\sqrt{fh}{\cal G}}{r}
+\frac{f'\sqrt{h}{\cal H}}{2\sqrt{f}}
 \right) }\,,
\ea
and 
\ba
e_4 &=&{\frac {1}{\phi'}}c_4'-{\frac {f'}{4f h \phi'^2}} 
\left( \sqrt{fh} {\cal H} \right)'
-{\frac {\sqrt {f}}{2\phi'^2\sqrt {h}r}}{\cal G}'
+{\frac {1}{h\phi' r^2} \left( {\frac {\phi''}{\phi'}}+\frac12 {\frac {h'}{h}} \right) }a_1
\notag\\
&&
+{\frac {\sqrt{f}}{8\sqrt{h}\phi'^2} 
\left[ {\frac { ( f' r-6 f ) f'}{f^2r}}
+\frac {h' ( f' r+4 f ) }{fhr}
-{\frac {4f ( 2 \phi'' h+h' \phi')}
{\phi' h^2r ( f' r-2 f ) }} \right] }{\cal H}
+{\frac {h'}{2h\phi'}}c_4
\notag\\
&&
+{\frac {f' hr-f}{2r^2\sqrt {f}{h}^{3/2}\phi'^2}}{\cal F}
+{\frac {\sqrt {f}}{2r\phi'^2{h}^{3/2}} 
\left[ {\frac {f ( 2 \phi'' h+h' \phi' ) }{h\phi'  ( f' r-2 f ) }}
+{\frac {2 f-f' hr}{2fr}} \right] }{\cal G}
\,, \label{e4}\\
d_3&=&
-{\frac {1}{r^2} \left( {\frac {2\phi''}{\phi'}}+{\frac {h'}{h}} \right) }a_1
+{\frac {f^{3/2}h^{1/2}}{ ( f' r-2 f ) \phi'} \left( 
{\frac {2\phi''}{h\phi' r}}
+ {\frac {{f'}^{2}}{f^2}}
- {\frac {f' h'}{fh}}
-{\frac {2f'}{fr}}
+{\frac {2h'}{hr}}
+ {\frac {h'}{h^2r}} \right) }{\cal H}
\notag\\
&&
+{\frac {\sqrt {f}}{\phi' \sqrt {h}r^2}}{\cal F}
-{\frac {{f}^{3/2}}{\sqrt {h} ( f' r-2 f ) \phi'} 
\left( {\frac {f'}{fr}}+{\frac {2\phi''}{\phi' r}}+{\frac {h'}{hr}}-\frac{2}{r^2} \right) }{\cal G}
\,.
\ea

\bibliographystyle{mybibstyle}
\bibliography{bib}

\end{document}